\documentclass[sigconf]{acmart}
\AtBeginDocument{%
  \providecommand\BibTeX{{%
    \normalfont B\kern-0.5em{\scshape i\kern-0.25em b}\kern-0.8em\TeX}}}

\copyrightyear{2023}
\acmYear{2023}
\setcopyright{rightsretained}
\acmConference[SIGGRAPH '23 Conference Proceedings]{Special Interest Group on Computer Graphics and Interactive Techniques Conference Conference Proceedings}{August 6--10, 2023}{Los Angeles, CA, USA}
\acmBooktitle{Special Interest Group on Computer Graphics and Interactive Techniques Conference Conference Proceedings (SIGGRAPH '23 Conference Proceedings), August 6--10, 2023, Los Angeles, CA, USA}
\acmDOI{10.1145/3588432.3591523}
\acmISBN{979-8-4007-0159-7/23/08}

\acmSubmissionID{489}

\usepackage{color}
\usepackage{float}
\usepackage{subcaption}
\usepackage{url}
\usepackage{wrapfig}
\usepackage{mathtools}
\usepackage[ruled]{algorithm2e} 

\usepackage{xcolor}

\SetCommentSty{mycommfont}
\usepackage{listings}

\def\textrightarrow{ -> }%

\definecolor{deepfuchsia}{rgb}{0.76, 0.33, 0.76}

\newcommand{\MB}[1]{#1}
\newcommand{\ME}[1]{#1}

\newcommand{\JS}[1]{#1}
\newcommand{\NGN}[1]{#1} 
\newcommand{\METD}[1]{#1} 

\def\xR{\mathbb{R}}
\newcommand{\tin}{\!\in\!}

\def\mM{M}
\def\mV{\mathcal{V}}
\def\mF{\mathcal{F}}
\newcommand{\nV}{n}
\newcommand{\nF}{m}
\newcommand{\mW}{W_D}
\newcommand{\mArea}{A}
\newcommand{\mgrad}{\nabla}

\newcommand{\mva}{A_\mathcal{V}}
\newcommand{\mfa}{A_\mathcal{F}}

\newcommand{\mgradd}{\text{G}}
\newcommand{\mdivd}{\text{D}}
\newcommand{\mvaM}{M_\mathcal{V}}
\newcommand{\mfaM}{M_\mathcal{F}}

\newcommand{\dist}{u}

\newcommand{\pmdist}{U}

\newcommand{\smW}{\alpha} 
\newcommand{\smE}{\mathcal{E}} 
\newcommand{\source}{x_0}
\newcommand{\sourceSet}{E}
\newcommand{\smWn}{\hat{\alpha}} 
\newcommand{\vfW}{\beta}
\newcommand{\vfWn}{\hat{\beta}}
\newcommand{\vf}{V}

\newcommand{\optW}{W}

\newcommand{\admmpen}{\rho} 
\newcommand{\admmlm}{y} 
\newcommand{\admmz}{z} 
\newcommand{\admmlmY}{Y} 
\newcommand{\admmZ}{Z} 
\newcommand{\admmlmu}{S} 
\newcommand{\admmlmm}{H} 
\newcommand{\admmlmn}{K} 
\newcommand{\admmzq}{Q} 
\newcommand{\admmzd}{U} 
\newcommand{\admmr}{R} 
\newcommand{\admmx}{X} 

\usepackage{enumitem,enumerate}
\setlist[itemize]{noitemsep, nolistsep, leftmargin=*}
\setlist[enumerate]{noitemsep, nolistsep, leftmargin=*}

\begin{document}

\title[A Convex Optimization Framework for Regularized Geodesic Distances]{A Convex Optimization Framework for \\ Regularized Geodesic Distances}


\author{Michal Edelstein}
\orcid{0000-0001-9126-1617}
\affiliation{%
    \institution{Technion - Israel Institute of Technology}
    \city{Haifa}
    \country{Israel}}
\email{smichale@cs.technion.ac.il}

\author{Nestor Guillen} 
\orcid{0000-0002-4940-7595}
\affiliation{%
    \institution{Texas State University}
    \city{San Marcos}
    \state{TX}
    \country{USA}}
\email{nestor@txstate.edu}

\author{Justin Solomon} 
\orcid{0000-0002-7701-7586}
\affiliation{%
    \institution{Massachusetts Institute of Technology (MIT)}
    \city{Cambridge}
    \state{MA}
    \country{USA}}
\email{jsolomon@mit.edu}

\author{Mirela Ben-Chen} 
\orcid{0000-0002-1732-2327}
\affiliation{%
    \institution{Technion - Israel Institute of Technology}
    \city{Haifa}
    \country{Israel}}
\email{mirela@cs.technion.ac.il}
 

\begin{abstract}
We propose a general convex optimization problem for computing \emph{regularized} geodesic distances.
 We show that under mild conditions on the regularizer the problem is well posed. We propose three different regularizers and provide analytical solutions in special cases, as well as corresponding efficient optimization algorithms. Additionally, we show how to generalize the approach to the \emph{all pairs} case by formulating the problem on the product manifold, which leads to \emph{symmetric} distances. Our regularized distances compare favorably to existing methods, in terms of robustness and ease of calibration.
\end{abstract}



\begin{teaserfigure}
  \includegraphics[width=\textwidth]{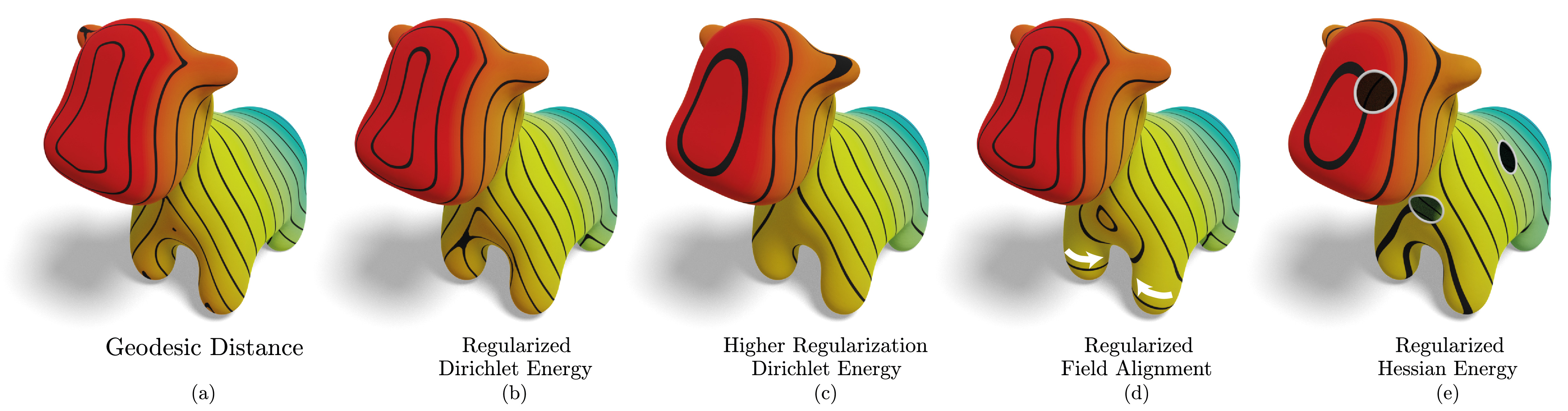}
  \caption{Geodesic distances (a) may not have desired properties such as smoothness. We present a general framework for \emph{regularized} geodesic distances. Shown here are three examples of regularizers: (b,c) smoothness, (d) alignment to a vector field, and (e) boundary invariance. }
  \Description{Teaser.}
  \label{fig:teaser}
\end{teaserfigure}


\maketitle

\section{Introduction}

Distance computation is a central task in shape analysis. Distances are required for many downstream geometry processing applications, including shape correspondence, shape descriptors and remeshing. In many cases, however, exact geodesic distances are not required, and a \emph{distance-like} function suffices. Moreover, it is often required to \emph{regularize} the distance-like function to improve performance of a downstream application. 

The geometry processing community has proposed myriad methods for computing geodesic distances~\cite{crane2020survey}, including some regularized distances~\cite{crane2013geodesics,solomon2014earth}. 
However, a unified framework, including a controlled and easily calibratable approach to regularization is still missing. 

We propose a flexible convex optimization framework for computing regularized geodesic distances. We show that under relatively mild conditions, our formulation has a minimizer that converges to the geodesic distance as the regularization weight vanishes. Furthermore, we propose a variety of regularizers, demonstrate their applicability, and provide corresponding efficient optimization algorithms. Finally, we formulate the \emph{all-pairs} problem, as a special case of our framework on the \emph{product manifold}. This formulation has the additional advantage that the resulting regularized distances are \emph{symmetric} with respect to swapping the source and target points. 

\subsection{Related Work}
The work on geodesic distances is vast, and a full review is out of scope. See the recent surveys~\cite{crane2020survey,peyre2010geodesic}.

\paragraph{Geodesic Distances} Some approaches (e.g., MMP~\cite{mitchell1987discrete}, MMP extension~\cite{surazhsky2005fast}, VTP~\cite{qin2016fast}, and others) compute the exact \emph{polyhedral} geodesic distance on a triangle mesh. Other methods, e.g., Fast Marching~\cite{kimmel1998computing}, take a variational approach and compute approximate geodesic distances. More recently, convex optimization approaches have been suggested for computing approximate geodesic distances~\cite{belyaev2015variational,belyaev2020admm} and for computing the cut locus~\cite{generau2022cut,generau2022numerical}. Our approach is also framed as a convex optimization problem, but incorporates an additional general regularization term. 

\paragraph{Regularization} Exact geodesic distances have some shortcomings in applications, e.g., the geodesic distance from a point $p$ is not smooth near the cut locus of $p$. The heat method~\cite{crane2013geodesics} computes smoothed geodesic distances, where the smoothing is controlled by a time parameter. The earth mover's distance (EMD)~\cite{solomon2014earth} can also be used to compute geodesic distances, optionally smoothed by projecting on a reduced spectral basis. Compared to the heat and EMD methods, our framework allows for a more direct control on the smoothness parameter. For triangle meshes, another option is to compute the graph-based distances on the graph of the triangulation with a Dirichlet or Hessian regularization~\cite{cao2020computing}. This approach is, however, triangulation dependent, and requires the use of $2$-ring neighborhoods for accurate results. Furthermore, we provide theoretical results that guarantee that our optimization problem is well-posed for general regularizers under some mild conditions, providing mathematical footing required for future work to design additional regularizers.

\subsection{Contributions}
Our main contributions are:
\begin{itemize}
    \item 
    A convex optimization problem for extracting regularized geodesic distances, with theoretical results for a general regularizer under some mild conditions. 
    \item Examples of regularizers with corresponding theoretical results and efficient optimization algorithms.
    \item The \emph{all-pairs} generalization, with a scalable optimization scheme.
\end{itemize}

\section{Background}
\subsection{Geodesic Distances by Convex Optimization}
Variational characterizations of the geodesic distance function are natural from several perspectives. From probability, they relate to large deviation estimates for the heat equation, as shown by Varadhan \shortcite{varadhan1967}. From a purely PDE perspective, they can be constructed as the largest viscosity subsolution to the Eikonal equation, using Ishii's extension of the Perron method to Hamilton-Jacobi equation \cite{ishii1987}. Recently, it was shown\NGN{~\cite{belyaev2015variational, belyaev2020admm}} how the geodesic distance $\dist$ on a domain $\Omega$ from a source point $\source$ 
can be computed by solving the convex optimization problem 
\begin{align}
\label{eq:gd}
    \begin{array}{rl}
    \textnormal{Minimize}_{\dist} & -\int_\Omega \dist(x)\;\text{dVol}(x) \\
    \textnormal{subject to} &  |\nabla \dist(x)|\leq 1 \text{ for all } x\in \Omega\setminus \{\source\}\\
	& \dist(\source) = 0.
    \end{array}
\end{align}
\NGN{As explained by Belyaev et al.~\shortcite{belyaev2020admm}, we can thus use convex optimization methods, e.g. ADMM, to approximate geodesic distances.} 

Intuitively, since the function $u$ is maximized, the gradient norm reaches the maximal allowed value, which is $1$. Therefore, while not constraining it directly, the solution will fulfill $|\nabla u|=1$ at every point in the domain and thus will be a geodesic distance. The big advantage of this formulation, as opposed to directly constraining $|\nabla u| = 1$, is that this optimization problem is \emph{convex}. \NGN{Furthermore, the point constraint $\dist(\source)=0$ may be relaxed to $\dist(\source)\leq 0$ without changing the solutions to the problem. To see why, note that if $\phi:M\to \mathbb{R}$ is such that $|\nabla \phi(x)|\leq 1$ for all $x \in \Omega\setminus\{\source\}$ and $\phi(\source)<0$, then the function $\tilde \phi := \phi-\phi(\source)$ will satisfy the two constraints in \eqref{eq:gd} and have a strictly smaller objective functional than $\phi$ since $\phi < \tilde \phi$ everywhere.}

\section{Regularized Geodesic Distances}
Given a compact surface $\mM$ and a closed set $\sourceSet \subset \mM$ (typically, $\sourceSet = \{\source\}$), our goal is to compute a function $\dist:\mM\to\xR$ which is ``as-geodesic-as-possible,'' but has some additional property. Depending on the application, one may require the function to be smooth,
or to be aligned to an input direction at some points on the surface. We assume that this additional information is encoded in a  \emph{regularizer} functional of the form
\begin{align}
  \smE(\dist) = \int_{\mM} F(\nabla \dist(x),x)\;\text{dVol}(x),
\label{eq:gen_reg_smooth}
\end{align}
where $F$ is convex in the first argument.

Generalizing Eq.\eqref{eq:gd}, we consider the following convex optimization problem 
\begin{align}
\label{eq:rgd}
    \boxed{
    \begin{array}{rl}
    \textnormal{Minimize}_{\dist} & \smW \smE(\dist) -\int_\mM \dist(x)\;\text{dVol}(x) \\
    \textnormal{subject to} &  |\nabla \dist(x)|\leq 1 \text{ for all } x\in \mM\setminus \sourceSet\\
	& \dist(x) \leq 0 \text{ for all } x \in \sourceSet.
    \end{array}
    }
\end{align}
with some $\smW > 0.$ 
We discuss various options for $\mathcal{E}$ in Sections \ref{sec:dir}, \ref{sec:objectives} and the supplemental.

\NGN{The problem \eqref{eq:rgd} has a long history in the case of a domain $\Omega \subset \mathbb{R}^d$ with $F(\nabla \dist(x),x) = |\nabla \dist(x)|^2$, which is known as the \emph{elastic-plastic torsion problem}. This is a free boundary problem, i.e., a PDE involving an interface, unknown a priori, across which the PDE's nature may change dramatically. In our case, this is reflected in two regions for the solution $\dist_\smW$, one where it solves a Poisson equation, and one where it solves the Eikonal equation. Refer to \cite{caffarelli1979free} and the book \cite{petrosyan2012regularity} for more background. For the Riemannian setting, this problem was first studied in \cite{generau2022cut}, discussed further below.}

Even under these general conditions (see the supplemental for detailed assumptions), we show that (a) the optimization problem has a  minimizer for every $\smW>0$ 
, (b) the minimizer is unique, and (c) they converge uniformly to the exact geodesic distance as $\smW\to 0$. We gather these results under the next two theorems.

\begin{theorem}\label{thm:regularized problem existence and uniqueness}
  There is a unique minimizer for problem \eqref{eq:rgd}.
\end{theorem}

\begin{theorem}\label{thm:regularizers uniform convergence}
  Let $\dist_\smW$ denote the minimizer to the optimization problem \eqref{eq:rgd}. Then, as $\smW \to 0$
  \begin{align*}
    \max\limits_{x\in M}|d(x,\sourceSet)-\dist_\smW(x)| \to 0,
  \end{align*}
  where $d(x,E)$ is the geodesic distance from $x$ to the set $E$.
\end{theorem}
\noindent The proofs of Theorem \ref{thm:regularized problem existence and uniqueness} and Theorem \ref{thm:regularizers uniform convergence}, are in Supp. 2 and 3.

\NGN{The unique minimizer $\dist_\smW$ provided by Thm. \ref{thm:regularized problem existence and uniqueness} is Lipschitz continuous by construction. 
In addition, it has two distinct regimes in the respective regions $\{|\nabla \dist_\smW| =1 \}$ and $\{|\nabla \dist_\smW|<1\}$.
For general second order elliptic regularizers, $\dist_{\smW}$ will be smooth in the interior of $\{|\nabla \dist_\smW|<1\}$, there $\dist_{\smW}$ will solve the unconstrained Euler-Lagrange equation corresponding to the objective functional in \eqref{eq:rgd}, which would be a nonlinear elliptic equation. Accordingly, standard elliptic theory guarantees that $\dist_\smW$ will be smooth in the region where the gradient constraint is not active. In the other region $\{|\nabla \dist_\smW| = 1 \}$ the function will solve the Eikonal equation in the viscosity sense. 

In the case $F = |\nabla \dist|^2$ (Sec.~\ref{sec:dir}) it was proved in \cite{generau2022cut} that for all $\smW < \smW_0$ ($\smW_0$ depending on the geometry of $\Omega$) the minimizer $\dist_\smW$ agrees with the geodesic distance function in $\{|\nabla \dist| =1 \}$. Therefore, $\dist_\smW$ coincides with the distance function everywhere save for a region around the cut locus of $\source$. In this region  $\dist_\smW$ solves the Poisson equation $\Delta \dist_\smW = -1/\smW$. As shown in \cite{generau2022cut}, as $\smW\to 0$, the open set $\{|\nabla \dist_\smW| <1 \}$ shrinks and converges to the cut locus. We expect the theorems in \cite{generau2022cut} to hold for general elliptic functionals (such as the $p$-Laplace equation), but this entails pointwise estimates of nonlinear elliptic equations on manifolds beyond the scope of this work. 

 Our regularizing functionals (Sec.~\ref{sec:dir},~\ref{sec:objectives}) correspond to elliptic energy functionals that promote smoothness and other desirable properties (non-negativity, symmetries) in the minimizer of \eqref{eq:rgd}. Accordingly, the significance of Thm. \ref{thm:regularizers uniform convergence} is in providing a smooth approximation to the geodesic distance function solution. Moreover, this approximation is in the $L^\infty$ metric, so the approximation error can be made small for all $x\in M$ provided $\smW$ is sufficiently small. }
 
\subsection{Dirichlet Regularizer}
\label{sec:dir}

A natural regularizer (also considered in~\cite{generau2022cut,generau2022numerical}) is the Dirichlet energy
\begin{align}
\label{eq:dirichlet}
  \smE_{\mathrm{Dir}}(\dist) & = \frac{1}{2}\int_{\mM}|\nabla \dist(x)|^2\;\text{dVol}(x).
\end{align}

\ME{First, we look at the simple example where the solution to problem \eqref{eq:rgd} using the Dirichlet energy \eqref{eq:dirichlet} is given by an explicit formula. We }
analyze the case where $\mM$ is the circle $\mathbb{S}^1$. Fix $\smW>0$. 
We parameterize $\mathbb{S}^1$ via the map $x\mapsto (\cos(x),\sin(x))$, i.e., by real numbers modulo $2\pi$. Moreover, we will use the group structure of $\mathbb{S}^1$, which is given by $\mathbb{R} / 2\pi \mathbb{Z}$. In this case the problem amounts to looking for a $2\pi$-periodic function $\dist(x)$ that minimizes
\begin{align*}
  \frac{\smW}{2}\int_{0}^{2\pi}|\dist'(x)|^2\;dx - \int_0^{2\pi}\dist(x)\;dx
\end{align*}
among all such $2\pi$-periodic functions satisfying the constraints
\begin{align*}
 \dist(0)\leq 0 \text{ and } | \dist'(x)|\leq 1 \text{ for all } x\in(0,2\pi).
\end{align*}
The minimizer $\dist(x)$ for this problem has a simple analytical expression, given as follows: First, given $x\in \mathbb{R}$ we set $\hat x = x \, \text{mod} \, 2\pi$. Then,
\begin{align}\label{eq:circle rgd example}
  \dist_\alpha(x) = \left \{ \begin{array}{ll}
  \hat x & \text { if } 0 \leq \hat x \leq L\\
  \pi-\frac{1}{2}\smW-\frac{1}{2\smW}(\hat x-\pi)^2 & \text{ if } L\leq \hat x \leq 2\pi-L\\
  2\pi-\hat x & \text{ if } \hat x \geq L
  \end{array} \right.
\end{align}
Here, $L = L(\smW)$ is defined by
\begin{align}\label{eq:circle example size smoothing region}
  L(\smW) =(\pi-\smW)_+.
\end{align}
This expression approximates the distance to the point corresponding to $x=0$. Observe that for $\smW>\pi$ the functions $\dist_\smW$ are all equal to $\dist_\pi$. In general, the solution $\dist_{\smW}$ has two regimes or regions, one region where it matches the geodesic distance function exactly, and one where it is solving Poisson's equation $\dist_{\smW}'' = -1/\smW$ and therefore matches a concave parabola, with the condition that $\dist_{\smW}$ is $C^1$ across these two regions. This is the standard condition for solutions to the obstacle problem (see \cite{petrosyan2012regularity}), which is intrinsically related to \eqref{eq:rgd} in this particular case (see Supplemental 1 for further discussion). 
\begin{wrapfigure}{r}{0.15\textwidth}
    \hspace*{-50pt}
    \vspace{-17pt}
    \begin{center}
        \hspace*{-10pt}
        \includegraphics[width=1.1\linewidth]{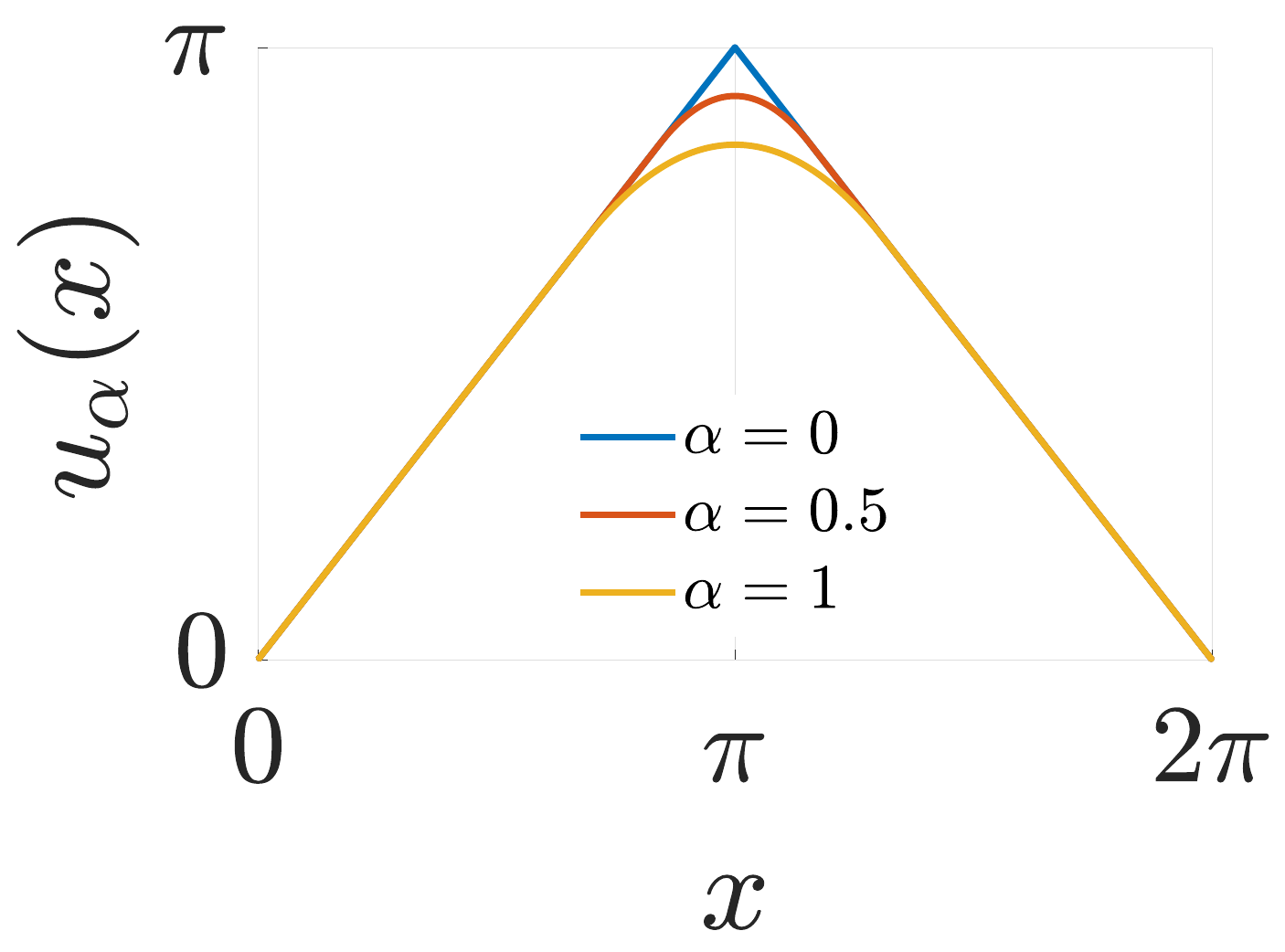}
    \end{center}
    \vspace{-15pt}
\end{wrapfigure}
\ME{The inset figure} shows the behavior of the function on the circle. Note the smoothing region, whose width depends on the smoothing parameter $\smW$ and matches \eqref{eq:circle rgd example}-\eqref{eq:circle example size smoothing region}.

Thanks to the group structure of $\mathbb{S}^1$ and the invariance of the problem under the group action 
(in other words, by symmetry), we obtain a corresponding formula when the source point is any point $y\in \mathbb{S}^1$. In particular, if $\dist_\alpha(\cdot,y)$ represents the solution to the problem with source at $y$, then
\begin{align}\label{eq:circle rgd invariance}
  \dist_\smW(x,y) = \dist_\smW(x-y),\;\;\forall\;x,y
\end{align}
We highlight a notable fact about these functions in a theorem.
\begin{theorem}\label{thm:circle example distance}
  For every $\smW>0$, the function $\dist_\smW(x,y)$ given by \eqref{eq:circle rgd example}-\eqref{eq:circle rgd invariance} defines a metric on $\mathbb{S}^1$. 
\end{theorem}

This theorem will be proved in the supplemental. For a general $M$, it is not clear whether one can expect $\dist_\smW(x,y)$ to be a metric. At the very least, it might be that Theorem \ref{thm:circle example distance} may generalize to other groups or homogeneous spaces. In Section \ref{sec:product formulation} we discuss an extension of problem \eqref{eq:rgd} to the product manifold $M\times M$ that treats all pairs $(x,y)$ at once, producing an approximation $\pmdist_{\smW}(x,y)$ that we can prove will be symmetric in $(x,y)$. We do not prove this general formulation has the triangle inequality but Figure \ref{fig:triangleinequality} provides some encouragement in that direction. 

\ME{Another simple example for which we can compute the analytical solution is the disk. We discuss it in the Supplemental, Section 1.}

\ME{For general triangle meshes, }
we provide the discrete formulation in Section \ref{sec:opt}.
Figure~\ref{fig:dirichlet_meshes} shows the behavior of our computed distance using the Dirichlet energy as $\smW$ changes. We plot the normalized error between $\dist_\smW$ and $\dist_0$, showing the solution converges smoothly towards $\dist_0$. We also show the result for three $\smW$ values. For each $\smW$, we see both the level sets of the distance function (right), and the norm of the gradient $| \mgrad \dist |$ (left). As $\smW$ increases the smoothing area around the cut locus becomes larger. Note that far from the cut locus the norm of the gradient is exactly one, showing that our regularized distance is exactly a geodesic distance function there. 

\section{Regularizers}
\label{sec:objectives}

In Section \ref{sec:dir} and Supplemental 1 we discussed two analytical examples using the Dirichlet energy \eqref{eq:dirichlet} as the regularizer. In this section we discuss other possible regularizers in various geometries (not using analytical formulas). The first of those functionals is included in the class \eqref{eq:gen_reg_smooth} covered by our theorems and is motivated by the question of alignment to a given vector field. The other one is a Hessian functional that falls outside the theorems in our work but for which we make several numerical experiments and which raises interesting theoretical questions. Lastly, we discuss one more example of a regularizer in Supplemental 6, with a non-quadratic regularizer that takes advantage of the general form of $\mathcal{E}$ in  \eqref{eq:rgd}.

\subsection{Vector Field Alignment}
\label{sec:objvfa}
\setcounter{figure}{1}
\begin{figure}
    \centering
    \includegraphics[width=\linewidth]{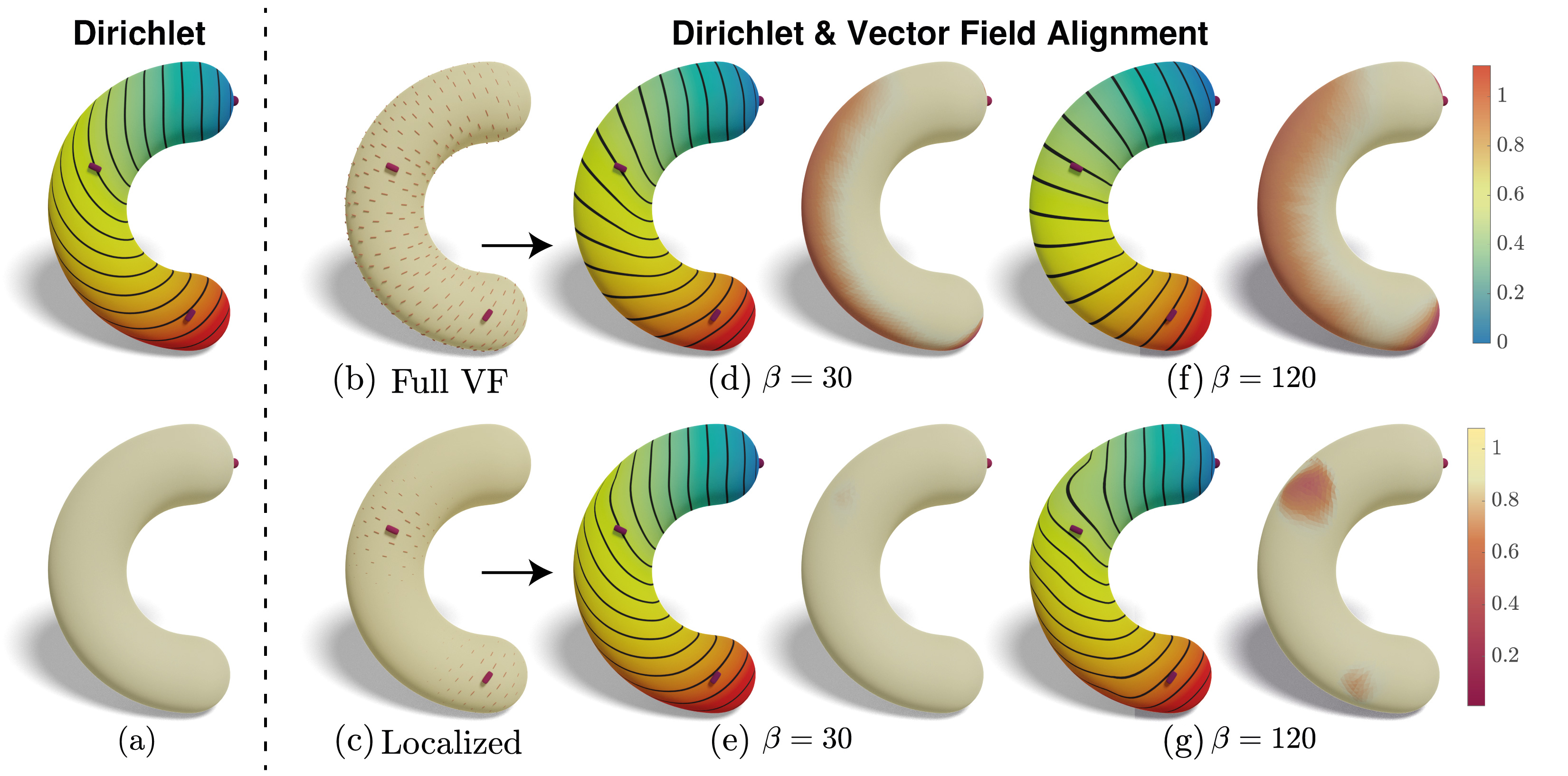}
    \caption{Vector field alignment regularization. (a) Dirichlet regularized distance. The two marked vector directions are not aligned with the regularized distance. (b) An interpolated and (c) localized vector field based on the two directions. (d-g) The corresponding regularized distance, using both Dirichlet and vector field alignment (Sec.~\ref{sec:objvfa}), using two regularization weights.}
    \label{fig:vfa}
\end{figure}

In addition to smoothing, one might want to align the isolines of the distance function to a given direction. We can align $\mgrad \dist$ with the line field $\vf (x)$ represented as a $3$D vector at each point $x$ using the following regularizer:
\begin{align}
\label{eq:reg_vfa_smooth}
  \smE(\dist) & = \frac{1}{2}\int_{\mM}|\nabla \dist(x)|^2+\vfW \langle V(x),\nabla \dist(x) \rangle ^2\;\text{dVol}(x),
\end{align}
Here, $(V(x),\nabla\dist(x))$ refers to the Riemmanian metric in $M$, which amounts to the usual inner product between vectors when $M$ is a surface in $\mathbb{R}^3$, for example. Note the additional parameter $\vfW$ that gives the relative weight between alignment to the vector field $V(x)$ versus general smoothness. 
This is equivalent to computing a distance function using an anisotropic smoothing term, where the anisotropic metric at each point on the surface is represented in $\mathbb{R}^3$ using the following matrix: $I + \vfW \hat{\vf}$ where $I$ is the identity matrix, and $\hat{\vf} = \vf \vf^T$. In terms of the Lagrangian $F(\xi,x)$ in the regularizer functional, this problem corresponds to choosing
\begin{align*}
  F(\xi,x) = |\xi|^2+ \beta \langle V(x),\xi \rangle ^2 = |A(x)\xi|^2
\end{align*}
where $A(x) = I+\vfW\hat{\vf}(x)$.

We allow the user to either provide a line field at each point, or specify a sparse set of directions.
If needed, we interpolate the sparse constraints to a smooth line field, as suggested by \citet[Section 5.5.4]{pluta2021ph}. Optionally, we scale the interpolated line field with a geodesic Gaussian (the geodesic distance is computed with our method without regularization).

Figure \ref{fig:vfa} shows the results. Starting from two vectors (a), we interpolate a line field (b), or a localized line field (c), and compute the resulting vector field aligned regularized distance for two regularization parameters $\beta$ (d-g). The gradient norm shows where the function deviates from being a geodesic distance as its isolines align to the prescribed directions.

\subsection{Hessian for Natural Boundary Conditions}

\ME{For the Dirichlet regularizer, if we do not impose any boundary conditions on the problem, the minimizer will have zero Neumann boundary values (sometimes called “natural boundary conditions” in FEM).} Recently, \citet{stein2020smoothness} suggested using the Hessian energy instead, given by
\begin{align}
\smE(\dist) & = \frac{1}{2}\int_{\mM} |\nabla^2 \dist(x)|^2\;\text{dVol}(x).
\label{eq:reg_hessian_smooth}
\end{align}
Here, we use the Frobenius norm of the matrix $\nabla^2\dist$ (accordingly, this norm relies on the Riemannian metric of $M$). 
This energy yields natural boundary conditions, making the result more robust to holes or mesh boundaries. 
\ME{In the Supplemental, Section 4, we show the analytical solution for the simple case of the circle.}

\begin{figure}
    \centering
    \includegraphics[width=\linewidth]{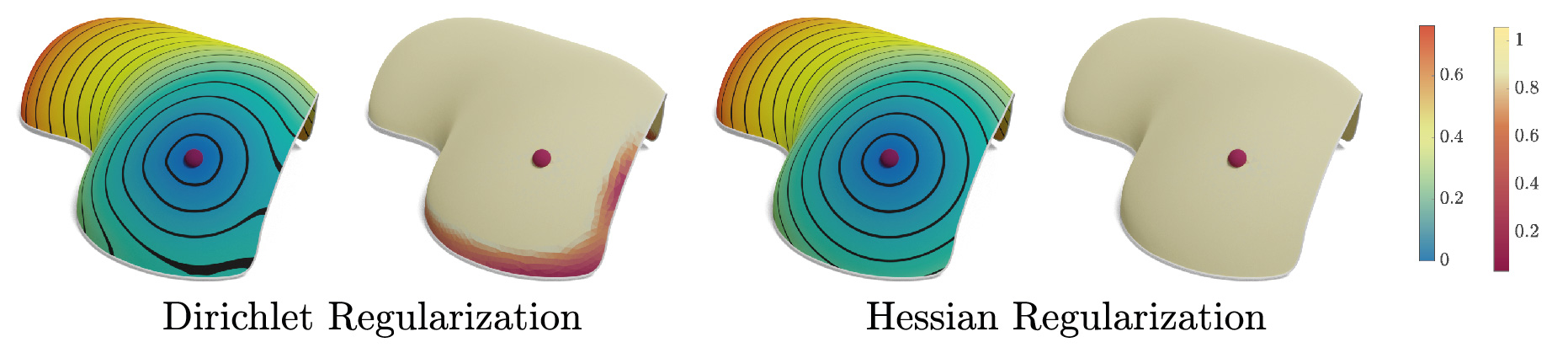}
    \caption{The distance computed using the Dirichlet energy regularizer (left) and the curved Hessian (right). Note the differences near the boundaries.}
    \label{fig:HNB}
\end{figure}

Figure \ref{fig:HNB} demonstrates this, using the Dirichlet energy (left) and the curved Hessian (right). For each method, the left image shows the level sets of the function, and the right image shows the norm of the gradient $| \mgrad \dist |$.  Note the difference near the mesh boundaries, where using the Dirichlet energy leads to zero Neumann conditions, meaning that the isolines are perpendicular to the boundary, and the distance is smoothed, whereas when using the Hessian energy the distance is unaffected by the boundary.

\section{Optimization via ADMM}
\label{sec:opt}

\subsection{Notation}

Discretely, we represent surfaces using triangle meshes $\mM=(\mV,\mF)$, where $\mV$ are the vertices, $\mF$ are the faces, and $\nV = | \mV |, \nF = | \mF |$.
We use a piecewise-linear discretization of functions on the mesh with one value per vertex; hence, functions are represented as vectors of length $\nV$. Vector fields are piecewise constant per face and can be represented in the trivial basis in $\mathbb{R}^3$ or in a local basis per face; we represent them using vectors of length $\mathbb{R}^{3\nF}$ or $\mathbb{R}^{2\nF}$, respectively.

Vertex and face areas are denoted by $\mva \in \mathbb{R}^{\nV}, \mfa \in \mathbb{R}^{\nF}$, where the area of a vertex is a third of the sum of the areas of its adjacent faces. The diagonal matrices $\mvaM  \in \mathbb{R}^{\nV \times \nV}$, $\mfaM  \in \mathbb{R}^{3\nF \times 3\nF}$ contain $\mva,\mfa$ on their corresponding diagonals (repeated $3$ times for $\mfa$). The total area of the mesh is $\mArea$. 
 We use standard differential operators~\cite[Chapter 3]{botsch2010polygon}. In particular, our formulation requires the cotangent Laplacian $\mW \in \mathbb{R}^{\nV \times \nV}$, the gradient $\mgradd \in \mathbb{R}^{3\nF \times \nV}$, and the divergence $\mdivd = \mgradd^T \mfaM \in \mathbb{R}^{\nV \times 3\nF}$.

\subsection{Optimization Problem}
In this setting, the optimization problem in Eq.~\eqref{eq:rgd}, becomes:
\begin{align}
    \label{eq:gen_reg_discrete}
    \begin{array}{rll}
    \textnormal{Minimize}_{\dist} & - \mva^T \dist + \smW F_\mM(\mgradd\dist) \\
    \textnormal{subject to} &  |(\mgradd \dist )_f |\leq 1 & \text{ for all } f\in \mF\\
	& \dist_{i} \leq 0 &\text{ for all } i \in \sourceSet,
    \end{array}
\end{align}
where $\sourceSet$ here is a subset of vertex indices where the distance should be $0$, and $F_\mM$ is a convex function that acts on the gradient of $\dist$. Note that this problem is convex whenever $F$ is convex, since the objective will be convex, the first constraint is a second-order cone constraint, and the second constraint is a linear inequality.

\subsection{Quadratic Objectives}
In practice, the objectives we consider are quadratic, leading to the following optimization problem
\begin{align}
    \label{eq:dg_d}
    \begin{array}{rll}
    \textnormal{Minimize}_{\dist} & - \mva^T \dist + \frac{\smW}{2} \dist^T \optW \dist \\
    \textnormal{subject to} &  |(\mgradd \dist )_f |\leq 1 & \text{ for all } f\in \mF\\
	& \dist_{i} \leq 0 & \text{ for all } i \in \sourceSet.
    \end{array}
\end{align}

Different functionals correspond to different weight matrices $\optW$. To use the \emph{Dirichlet energy} in Eq.~\eqref{eq:dirichlet}, we set $\optW$ to the cotangent Laplacian  matrix $\mW$. For the \emph{vector field alignment} objective in Eq.~\eqref{eq:reg_vfa_smooth} we construct the anisotropic smoothing matrix $\optW_V=\mdivd(I+\vfW\hat{\vf}) \mgradd$, where $I \in \mathbb{R}^{3\nF \times 3\nF}$, and $\hat{\vf} \in \mathbb{R}^{3\nF \times 3\nF}$ is block diagonal, with the $3\times3$ block of face $f\tin\mF$ given by $V_f V_f^T$ (see also Section~\ref{sec:objvfa}). Finally, to use the \emph{Hessian} regularizer in Eq. ~\eqref{eq:reg_hessian_smooth} we take $\optW$ to be the curved hessian matrix~ \cite{stein2020smoothness}, denoted by $\optW_H$.

\subsection{Efficient Optimization Algorithm}\label{sec:admm_optim}
We derive an alternating direction method of multipliers (ADMM) algorithm~\cite{boyd2011distributed} to solve the optimization problem in Eq.~\eqref{eq:dg_d} efficiently. We reformulate the optimization problem, adding an auxiliary variable $\admmz \in \mathbb{R}^{3\nF}$ representing the gradient of the distance function $\mgradd \dist$. This leads to:

\begin{align*}
    \begin{array}{rl}
    \textnormal{Minimize}_{\dist} & -\mva^T \dist + \frac{1}{2}\smW \dist^T \optW \dist + \displaystyle \sum_{f \in \mF} \chi(|\admmz_f| \le 1) \\
    \textnormal{subject to} &  (\mgradd \dist)_f =\admmz_f \qquad \text{ for all } f\in \mF\\
	& \dist_{i} \leq 0  \qquad \qquad \text{ for all } i \in \sourceSet,
    \end{array}
\end{align*}
where $\chi(\cdot )$ is the indicator function, i.e., $\chi(|\admmz_f| \le 1) = \infty$ if $|\admmz_f| > 1$ and $0$ otherwise.

The corresponding augmented Lagrangian is:

\begin{align*}
    \begin{array}{rl}
    L(\dist,\admmlm,\admmz) &=   -\mva^T \dist + \frac{\smW}{2} \dist^T \optW \dist + \displaystyle \sum_{f \in \mF} \chi(|\admmz_f| \le 1) + \\
    & \displaystyle \sum_{f \in \mF} a_f \admmlm_f^T ((\mgradd \dist)_f - \admmz_f) + \frac{\admmpen \sqrt{\mArea}}{2} \displaystyle \sum_{f \in \mF} a_f |(\mgradd \dist)_f - \admmz_f|^2,
    \end{array}
\end{align*}
where $a_f$ is the area of the face $f$, $\admmpen \in \mathbb{R}$ is the penalty parameter, and $\admmlm \in \mathbb{R}^{3\nF}$ is the dual variable or lagrange multiplier.

The ADMM algorithm consists of iteratively repeating three steps \cite[Section 3]{boyd2011distributed}. First, we perform $\dist$-minimization, then $\admmz$-minimization, and finally the dual variable, $\admmlm$, is updated. The full derivation of the three steps appears in Supplemental 8, and the resulting algorithm in Algorithm \ref{alg:admm}.

\paragraph{Algorithm details} Note that the first step, the $\dist$-minimization, includes solving a linear system with a fixed coefficient matrix, which is pre-factored and used for all the ADMM iterations, as well as all distance computations. To enforce the constraint $\dist_{\sourceSet} \leq 0$, we eliminate the relevant columns from the linear system and solve for $\dist_{i} $ for all $i\in \mV\setminus \{\sourceSet\}$. We project intermediate $\admmz$ values to the unit ball, i.e. $\text{Proj}(z_f\!\in\!\xR^3,\mathbb{B}^3)$ is equal to $z_f/|z_f|$ if $|z_f|>1$, and $z_f$ otherwise. We use the stopping criteria suggested by Boyd et al.~\shortcite[Section 3.3.1]{boyd2011distributed}, formulated for our problem. See Supplemental 8 for details.

\begin{algorithm}[b]
    \smaller
    \caption{ADMM.\vspace{-.2in}}
    \SetKwInOut{Input}{input}\SetKwInOut{Output}{output}
    
    \Input{$\mM, \smW, \optW, \sourceSet$}
    \Output{$\dist \in \mathbb{R}^{\nV}$ - distance to $\sourceSet$}
   
    initialize $\admmpen \in \mathbb{R}$ \tcp*[r]{penalty parameter}
    
    \hskip3.7em $\admmz \gets \mathbf{0}^{3\nF}$ \tcp*[r]{auxiliary variable, $\mgradd \dist = \admmz$}
    \hskip3.7em $\admmlm \gets \mathbf{0}^{3\nF}$ \tcp*[r]{dual variable}
    \hskip3.7em $\admmpen \gets \admmpen \sqrt{\mArea}$
    
        \While(\tcp*[f]{See Supp. 8}){algorithm did not converge}{   
        
        $\text{solve}\,\big(\smW \optW+\admmpen \mW\big) \dist = \mva - \mdivd \admmlm + \admmpen \mdivd \admmz \quad \text{s.t.} \quad \dist_\sourceSet = 0$   \\
        
        $\admmz_f \leftarrow  \text{Proj} (\frac{1}{\admmpen } \admmlm_f + (\mgradd \dist)_f, \mathbb{B}^3)$
        \text{ for all } $f\in \mF $\\
        
         $\admmlm \leftarrow \admmlm + \admmpen  ( \mgradd \dist - \admmz )$ \\ 
    }
    \label{alg:admm}
\end{algorithm}

\paragraph{Efficiency}
We compare our approach with a solution implemented using commercial software, 
i.e., CVX~\cite{cvx,gb08} with the MOSEK solver~\cite{mosek}. Table~\ref{tab:timings} provides a comparison of the running times, measured on a desktop machine with an Intel Core i9. For the optimization using CVX and MOSEK, we report both the total running times and the solver running times. Our optimization scheme yields at least an order of magnitude improvement.

\begin{figure}
    \centering
    \includegraphics[width=\linewidth]{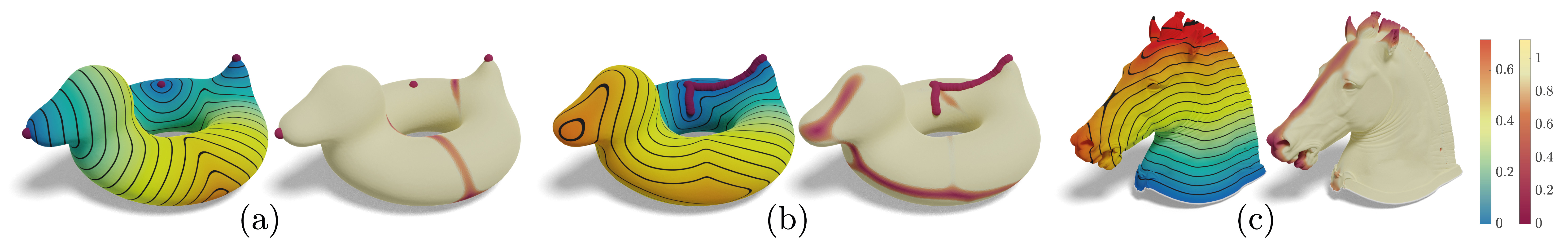}
    \caption{Distance to multiple sources: (a) $3$ points, (b) a vertex sampling of a path, and (c) the boundary. We show the distance and the gradient norm.}
    \label{fig:ssms}
\end{figure}

\begin{figure}[H]
    \centering
    \includegraphics[width=\columnwidth]{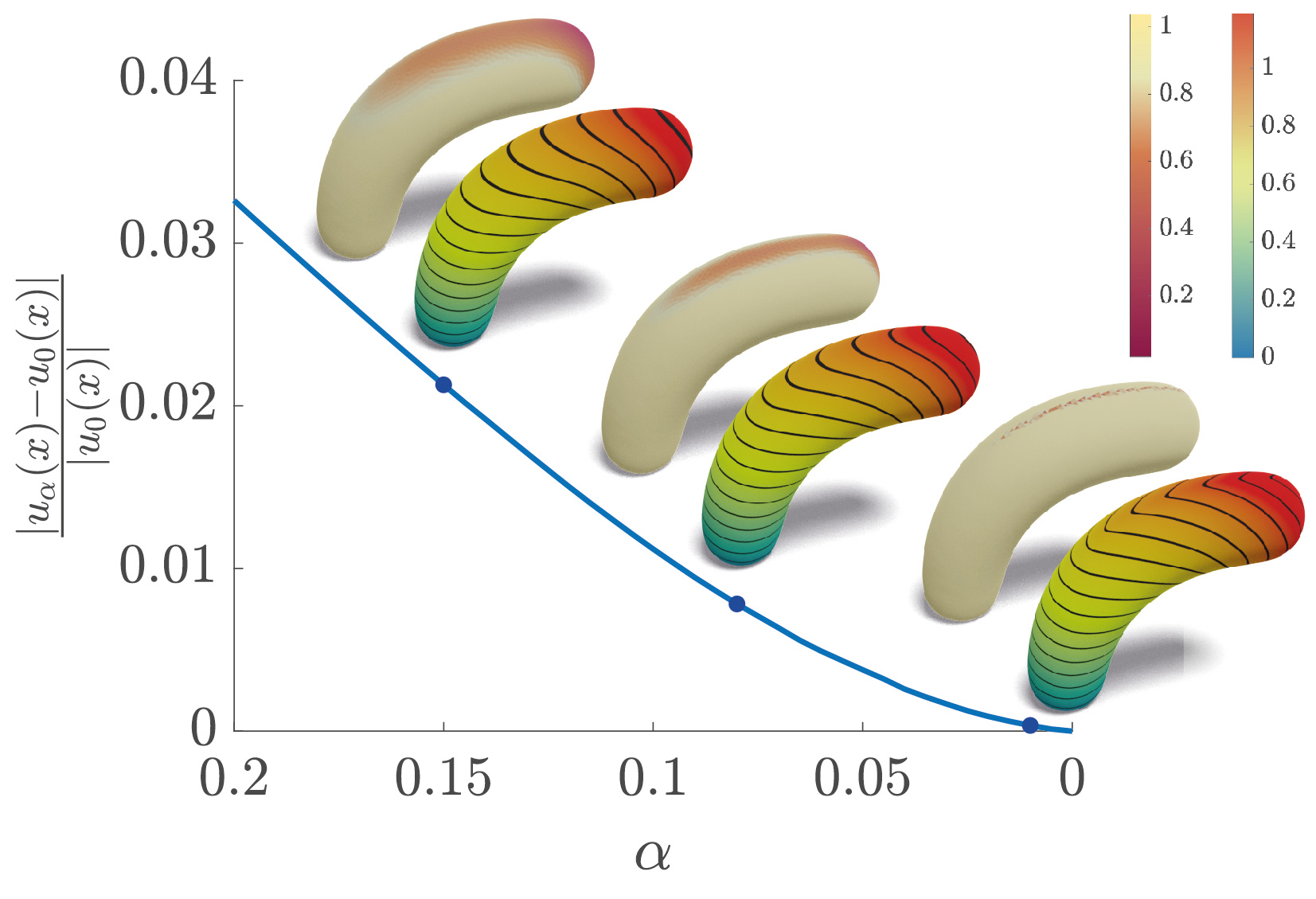}
    \caption{A numerical solution on triangle meshes. The distance converges towards $\dist_0$ as $\smW$ approaches $0$.  Note the different smoothing regions, whose width depends on $\smW$.}
    \label{fig:dirichlet_meshes} 
\end{figure}

Figure~\ref{fig:ssms} shows our result for multiple sources: three isolated points, a vertex sampling of a path, and the boundary. Additional results are shown in Section~\ref{sec:results} \MB{and in the Supplemental}.

\begin{table}
    \centering
    {\smaller
    \caption{Running times for computing the distance from a single source.} 
    \label{tab:timings}
        \begin{tabular}{  c |  c| c |c c  }             
            Model & $| \mF |$ & ADMM & \multicolumn{2}{c}{CVX (sec)} \\
            & & (sec)  & Total & MOSEK \\
            \hline
            Pipe, Fig~\ref{fig:dirichlet_meshes} & 10K & 0.075 & 1.16 & 0.36 \\ 
            Moai, Fig~\ref{fig:cmp} & 43K & 1.01 & 5.93 & 2.25\\ 
            Armadilo, Supp. 10 & 346K & 1.89 & 37.3733 & 21.88 \\ 
            Gardet, Supp. 10 & 989K & 5.22 & 132.91 & 87.83  \\
            Dragon, Supp. 10 & 2349K & 7.66 & 347.73 & 230.02 \\  
            Sea star, Supp. 10 & 3500K &  11.16 & 572.56 & 380.36 \\ 
    \end{tabular}
    }
    \vspace{-.1in}
\end{table}

\section{Symmetric All-Pairs Formulation}\label{sec:product formulation}

In Section \ref{sec:dir} we observed how the problem \eqref{eq:rgd} produces a new metric on $\mathbb{S}^1$. It is not clear if problem \eqref{eq:rgd} produces such a result in general. Part of the problem is how \eqref{eq:rgd} treats the $x$ and the source $y$ (if $\sourceSet = \{y\}$) differently. If one is interested in approximating the full geodesic distance function $d(x,y)$ it is of interest to have a formulation that works directly in all of $M\times M$ and which is naturally symmetric. This leads to the following variation on problem \eqref{eq:rgd}. 

Consider the manifold $M\times M$ with the product metric inherited from $M$. Naturally, given a function $\pmdist: M\times M \to \mathbb{R}$ we can fix $y\in M$ and consider the function $x\mapsto \pmdist(x,y)$ or fix $x\in M$ and consider the function $y\mapsto \pmdist(x,y)$. We define $\nabla_1 \pmdist(x,y)$ and $\nabla_2 \pmdist(x,y)$
to be the respective gradients for these functions in $M$. Equivalently, from the decomposition $T(M\times M)_{(x,y)} = (TM)_x\oplus (TM)_y$ we see that $\nabla_{M\times M} \pmdist(x,y) = (\nabla_1 \pmdist,\nabla_2 \pmdist)$.

With this notation, consider the minimization problem
\begin{align}\label{eq:rgdpm}
    \boxed{
    \begin{array}{rl}
    \textnormal{Minimize}_{\pmdist} & \smW \smE_{M\times M}(\pmdist) -\int_{\mM \times \mM} \pmdist(x,y)\;\text{dVol}(x,y) \\
    \textnormal{subject to} &  |\nabla_1 \pmdist(x,y)|\leq 1 \text{ in } \{ (x,y) \mid x\neq y\} \\
    & |\nabla_2 \pmdist(x,y)| \leq 1 \text{ in } \{ (x,y) \mid x \neq y\}\\
	& \pmdist(x,y) \leq 0 \text{ on } \{ (x,y) \mid x=y\}
    \end{array}
    }
\end{align}
Here, we are focusing on the Dirichlet energy functional
\begin{align*}
  \smE_{\mM\times \mM}(\pmdist) := \frac{1}{2}\int_{\mM\times \mM}|\nabla_1 \pmdist(x,y)|^2+|\nabla_2 \pmdist(x,y)|^2\;\text{dVol}(x,y)
\end{align*}
The optimization problem \eqref{eq:rgdpm} is a natural extension of \eqref{eq:rgd} if one is interested in the full geodesic distance function. Indeed, for $\smW = 0$, problem \eqref{eq:rgdpm} has only one minimizer, the geodesic distance function. 

\subsection{Theoretical Results}

Our discussion suggests that the solutions to \eqref{eq:rgdpm}, to the extent they exist, should converge to the geodesic distance as $\smW \to 0$. The next two theorems, counterparts to Theorems \ref{thm:regularized problem existence and uniqueness} and \ref{thm:regularizers uniform convergence}, address this. 

\begin{theorem}\label{thm:product formulation existence uniqueness}
  There is a unique minimizer for problem \eqref{eq:rgdpm}.    
\end{theorem}
\noindent (See Supplemental 5 for a proof).

We will denote the unique minimizer for this problem by $\pmdist_\smW(x,y)$. The motivation for \eqref{eq:rgdpm} was in part finding a way to guarantee the symmetry of the resulting regularization, and so we have the following theorems, proved in  Supplemental 5.

\begin{theorem}\label{lem:product formulation symmetry}
  The function $\pmdist_\smW(x,y)$ is symmetric in $x$ and $y$.
\end{theorem}

\begin{theorem}\label{thm:product formulation convergence}
   As $\smW\to 0$, we have
   \begin{align*}
     \|d(x,y)-\pmdist_\smW(x,y)\|_{L^\infty(\mM\times \mM)} \to 0.
   \end{align*}
\end{theorem}

Analogously to Theorem \ref{thm:regularizers uniform convergence}, this last theorem guarantees the functions $\pmdist_\smW$ provide a uniform approximation to the full geodesic distance $d(x,y)$ provided $\smW$ is chosen adequately. 

\subsection{Scalable Optimization}
\label{sec:optpm}

Discretely, we represent $\pmdist$ as an $\nV \times \nV$ matrix. We also express $\nabla_1 \pmdist(x,y), \nabla_2 \pmdist(x,y)$ as gradients over the row and columns of $\pmdist$, i.e., $\mgradd \pmdist$ and $\mgradd \pmdist^T$. 
The optimization problem in Eq.~\eqref{eq:rgdpm} becomes:

\begin{equation}
\label{eq:dgdpm}
    \begin{array}{rl}
    \textnormal{Minimize}_{\pmdist} & 
     - \mva^T \pmdist \mva \,+ \\
     & \frac{1}{2}\smW \, \text{Tr}\left(\mvaM \big(\pmdist^T \mW \pmdist + 
       \pmdist \mW \pmdist^T \big) \right) \\
    \textnormal{subject to} 
    & |(\mgrad \pmdist_{(i,\cdot)} )_f |\leq 1 \quad \text{ for all } f\in \mF, i\in \mV\\
    & |(\mgrad \pmdist_{(\cdot,j)} )_f |\leq 1 \quad \text{ for all } f\in \mF, j\in \mV\\
	& \pmdist_{i,i} \leq 0 \,\,\, \qquad \qquad \text{ for all } i\in \mV,
    \end{array}
\end{equation}
where $X_{i,j}$ denotes the $(i,j)$-th element of a matrix $X$,  $X_{(i,\cdot)}$ denotes the $i$-th row, and $X_{(\cdot,j)}$ the $j$-th column. 

The complexity here is significantly higher than computing the distance of all points to a closed set. 
\ME{A naive formulation of the ADMM for this problem leads to a per-iteration linear solve with a system matrix of size $\nV^2\times\nV^2$. To reduce it to $\nV$ solves with a system matrix of size $\nV\times\nV$, we }
derive a second ADMM algorithm, Alg. \ref{alg:admmpm} (see Supp), scalable to larger meshes. 
\ME{The symmetric formulation in Equation \eqref{eq:rgdpm} arises naturally in this derivation.}

\begin{algorithm}[b]
\smaller
    \SetKwInOut{Input}{input}\SetKwInOut{Output}{output}
    \Input{$\mM, \smW$}
    \Output{$\admmzd \in \mathbb{R}^{\nV \times \nV}$ \tcp*[r]{dual consensus variable}}
   
    initialize $\admmpen_1, \admmpen_2 \in \mathbb{R}$ \tcp*[r]{penalty parameters} 
    \hskip1em $\admmZ, \admmzq \gets \mathbf{0}^{3\nF \times \nV}$ \tcp*[r]{auxiliary variables $\mgradd \admmx = \admmZ$, $\mgradd \admmr = \admmzq $}
    \hskip1em $\admmlmY, \admmlmu \gets \mathbf{0}^{3\nF \times \nV}$ \tcp*[r]{dual variables}
    \hskip1em $\admmlmm, \admmlmn \gets \mathbf{0}^{\nV \times \nV}$ \tcp*[r]{dual consensus variables}
    \hskip1em $\admmpen_1 \gets \admmpen_1 \sqrt{\mArea}, \quad \admmpen_2 \gets \admmpen_2 \sqrt{\mArea^{-1}}$\\
    \hskip1em $\optW_{P} \gets (\smW + \admmpen_1)\mW + \admmpen_2 \mvaM $, \quad $M_{P} \gets \frac{1}{2} \mva \mva^T \mvaM^{-1} $
    
    \While(\tcp*[f]{See Supp. 9}){algorithm did not converge}{  
    
        solve for $\admmx$ \\
        \quad $ \optW_P \admmx = $
        $ M_P - \mdivd \admmlmY + \admmpen_1 \mdivd \admmZ  - \mvaM \admmlmm + \admmpen_2  \mvaM \admmzd$ \\
        solve for $\admmr$ \\
        \quad $ \optW_P \admmr = $ 
         $ M_P - \mdivd \admmlmu + \admmpen_1  \mdivd \admmzq  - \mvaM \admmlmn + \admmpen_2  \mvaM \admmzd^{T}$ \\  

        $(\admmZ_{(\cdot,i)})_f \leftarrow$
        \\ \quad \text{Proj} $\left( \frac{1}{\admmpen_1 } (\admmlmY_{(\cdot,i)})_f + (\mgradd \admmx_{(\cdot,i)})_f,\mathbb{B}^3 \right)  \text{ for all } i \!\in \!\mV, f\!\in\!\mF $\\
                
        $(\admmzq_{(\cdot,i)})_f \leftarrow$
        \\ \quad \text{Proj} $\left( \frac{1}{\admmpen_1} (\admmlmu_{(\cdot,i)})_f + (\mgradd 
        \admmr_{(\cdot,i)})_f,\mathbb{B}^3 \right) \text{ for all } i\! \in\! \mV, f\!\in\! \mF $\\

        $\admmzd = \text{max}\left(\frac{\admmlmm+\admmlmn^T}{2\admmpen_2 } + \frac{\admmx+\admmr^T}{2}, 0\right)$ ;  \quad    $\admmzd_{i,i} = 0$ \text{ for all } $i \in \mV$\\

        $\admmlmY \leftarrow \admmlmY + \admmpen_1  ( \mgradd \admmx - \admmZ );$ \quad 
        $\admmlmu \leftarrow \admmlmu + \admmpen_1  ( \mgradd \admmr - \admmzq )$\\
        $\admmlmm \leftarrow \admmlmm + \admmpen_2  ( \admmx - \admmzd );$ \quad 
        $\admmlmn \leftarrow \admmlmn + \admmpen_2  ( \admmr - \admmzd^T )$\\
    }  
    \caption{Symmetric All-Pairs ADMM.\vspace{-.2in}}
    \label{alg:admmpm}
\end{algorithm}

\begin{figure}[t]
    \centering
    \includegraphics[width=\columnwidth]{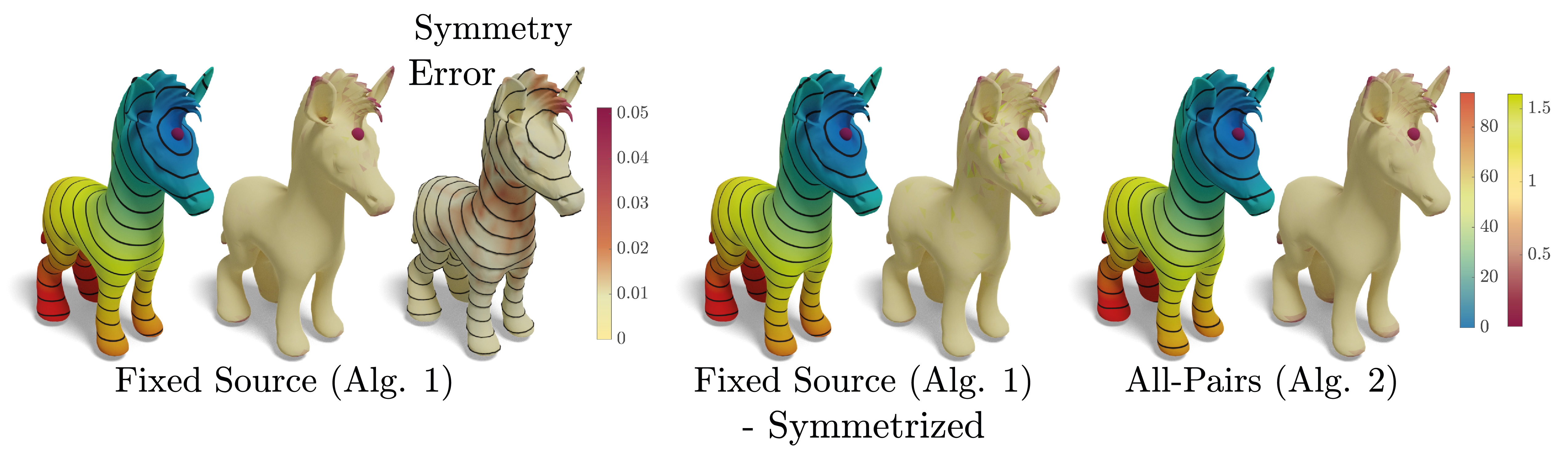}
    \caption{The all-pairs formulation, Alg.~\ref{alg:admmpm} (right), vs the fixed source formulation, Alg.~\ref{alg:admm} (left), and its symmetrized version (center). See the text for details.}
    \label{fig:pm} 
\end{figure}

Figure~\ref{fig:pm} shows an example of distances computed using this approach. The fixed source formulation (Alg.~\ref{alg:admm}) (left) is not symmetric. We can symmetrize the distance matrix (center), but this leads to visible noise in the gradient norm. The all-pairs formulation (Alg.~\ref{alg:admmpm}) is both symmetric and has a smooth gradient norm.

\section{Experimental Results}

\label{sec:results}
\subsection{Scale-Invariant Parameters}
\label{sec:scaleinv}

\begin{figure}
    \centering
    \includegraphics[width=\columnwidth]{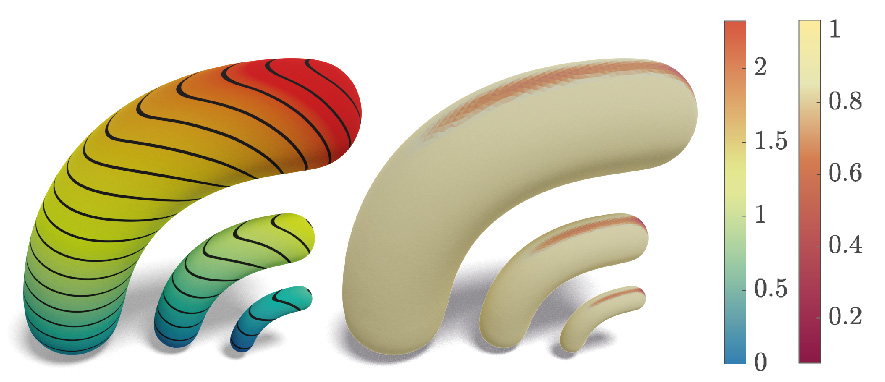}
    \caption{Scale invariance. While the distances are different between the uniformly scaled models, the area of the smoothing (where the norm of the gradient is not $1$) is similar for all meshes. See the text for details.}
    \label{fig:scaleinv} 
\end{figure}

The parameter $\smW$ controls the size of the smoothing area. Therefore, scaling the mesh requires changing its value.  To avoid that, and enable more intuitive control of the smoothing area, we define a scale-invariant smoothing parameter $\smWn$ that is independent of the mesh area or resolution. For the Dirichlet and vector field alignment energies, we achieve that by setting $\smW = \smWn \sqrt{\mArea}$. For the Hessian energy, we set $\smW = \smWn \sqrt{\mArea^3}$. We note that the parameter $\vfW$ is already scale-invariant, i.e., $\vfW = \vfWn$. For our ADMM algorithms (Sec. \ref{sec:opt}, \ref{sec:optpm}) to be scale-invariant, we normalize the penalty variables, residual and feasibility tolerances. 
Figure~\ref{fig:scaleinv} demonstrates this. We uniformly rescale an input mesh, and use the same smoothing parameter $\hat{\smW}$. Note that while the distances are different between the meshes, the scale of the smoothed region, i.e., the area where the norm of the gradient is not $1$, is similar. For all our experiments we use the scale invariant formulation, unless stated otherwise.

\subsection{Comparison}
\begin{figure}
    \centering
    \includegraphics[width=\columnwidth]{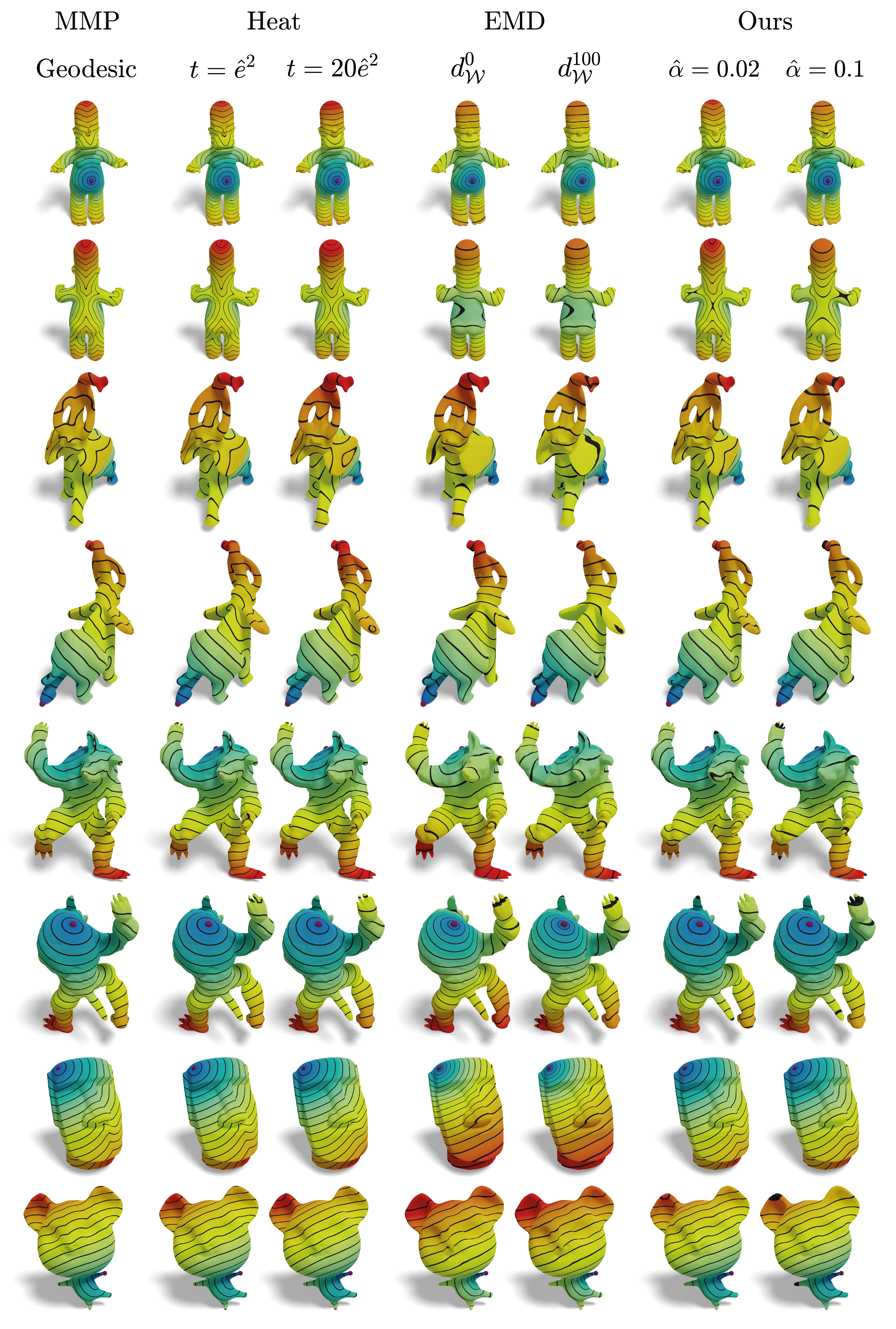}
    \caption{A qualitative comparison between our Dirichlet regularized distances, Heat method, and EMD, with two choices of smoothing parameter per method. See the text for details.}
    \label{fig:cmp} 
\end{figure}

In Fig.~\ref{fig:cmp} we compare our Dirichlet regularized distances to ``Geodesics in Heat''~\cite{crane2013geodesics} and regularized EMD~\cite{solomon2014earth}, with two smoothing parameters for each. In addition, we show the exact geodesics computed using MMP~\cite{mitchell1987discrete} for reference. 
Note that while all approaches lead to a smoother solution compared to the exact geodesics, our approach is more stable, in the sense that the same scale of regularization is observed on all meshes, for the same parameters. Thus, we conjecture that for our approach the regularization parameter is easier to tune.

\ME{Table \ref{tab:compte} compares the running times, and the maximal error w.r.t the MMP distance (as a $\%$ of the maximal distance). The distances are computed with Geometry Central \cite{geometrycentral} for the heat method and MMP, and with a Matlab implementation of our ADMM Algorithm \ref{alg:admm}. Note that both our method and the heat method have comparable errors, and for both smoother solutions have larger errors. A timing comparison for the all-pairs case is in the Supp.}

We additionally show in the supplemental material a comparison of the representation error of the different approaches in a reduced basis (providing a quantitative measure of smoothness).

\begin{table*}
    \centering
    {\smaller
    \caption{\ME{Comparison of run-times (T) and the maximal error ($\epsilon$) of the computed distance (in $\%$ of the maximal distance) for the models in Figure \ref{fig:cmp}.}}
    \label{tab:compte}
        \begin{tabular}{  c | c | c | c c | c c | c | c | c c | c c }             
            Model & $| \mF |$ &  MMP &  \multicolumn{2}{c|}{Heat $t=\hat{e}^2$} &  \multicolumn{2}{c|}{Heat $t=20\hat{e}^2$} & EMD $d_\mathcal{W}^0$ & EMD $d_\mathcal{W}^{100}$ &  \multicolumn{2}{c|}{Ours $\smWn=0.02$} &  \multicolumn{2}{c}{Ours $\smWn=0.1$}  \\
            
             &  & T (sec)  & T (sec) & $\epsilon$ (\%) & T (sec) & $\epsilon$ (\%) & $\epsilon$ (\%) & $\epsilon$ (\%) & T (sec) & $\epsilon$ (\%) & T (sec) & $\epsilon$ (\%) \\
            \hline

            Homer & 23K & 0.255 & 0.031 & 2.47 & 0.030 & 7.76 & 30.33 & 30.31 & 0.3664 & 3.40 & 0.1503 & 11.90 \\
            
            Elephant & 10K & 0.061 & 0.011 & 4.84 & 0.011 & 11.36 & 11.36 & 15.82 & 0.191 & 2.00 & 0.138 & 5.72 \\
            
            Armadilo & 29K & 0.279 & 0.041 & 2.83 & 0.041 & 9.33 & 20.55 & 14.82 & 0.454 & 2.35 & 0.192 & 6.78 \\
            
            Moai & 43K & 1.133 & 0.101 & 1.71 & 0.102 & 5.34 & 26.32 & 26.43 & 0.683 & 1.84 & 0.390 & 8.60 \\
            
            Koala & 9K & 0.063 & 0.013 & 2.30 & 0.010 & 7.17 & 55.26 & 41.15 & 0.124 & 2.35 & 0.059 & 8.64 \\

    \end{tabular}
    \vspace{-.1in}
    }
\end{table*}

\subsection{Robustness}
\begin{figure}
    \centering
    \includegraphics[width=\columnwidth]{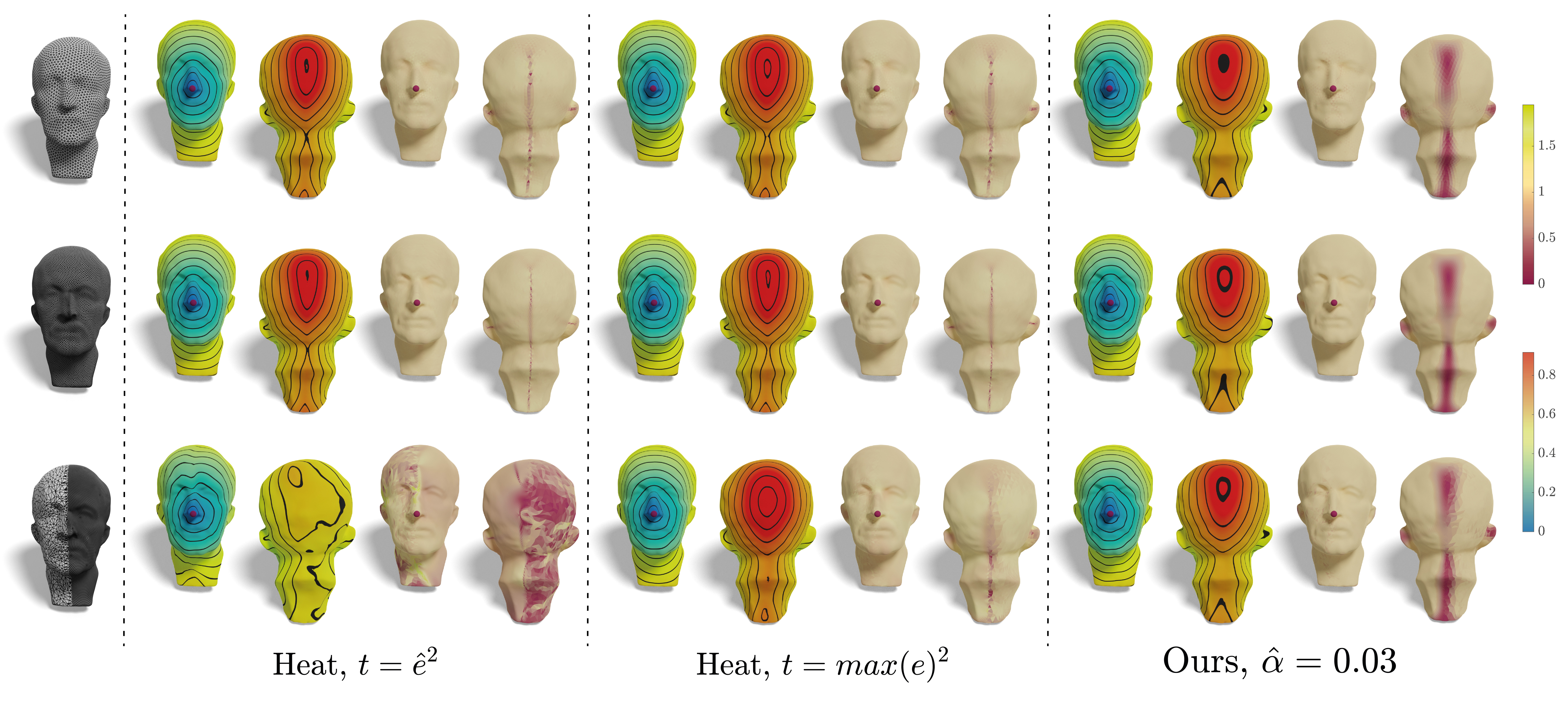}
    \caption{Our results v.s. the heat method for different remeshings, we show the distance function and the gradient norm. Note that our approach leads to distances which are very similar for the three meshes, \emph{with the same smoothing parameter $\hat\smW$}. See the text for details.  }
    \label{fig:meshing} 
\end{figure}

\emph{Meshing.}
We demonstrate that our method is invariant to meshing, and is applicable to non-uniform meshing \emph{without} modifying $\smW$. Fig.~\ref{fig:meshing} compares our result with the heat method, for $3$ remeshings of the same shape.
Note that for the heat method with the default smoothing parameter (left), the half-half mesh fails. This is remedied by using a different parameter (center), however there are still differences between the different meshing (note especially the gradient norm). Using our approach (right) we get very similar distance functions and gradient norm for all $3$ meshings. \MB{Fig. 5 in the supplemental shows additional results with bad triangulations.}

\begin{figure}
    \centering
    \includegraphics[width=\columnwidth]{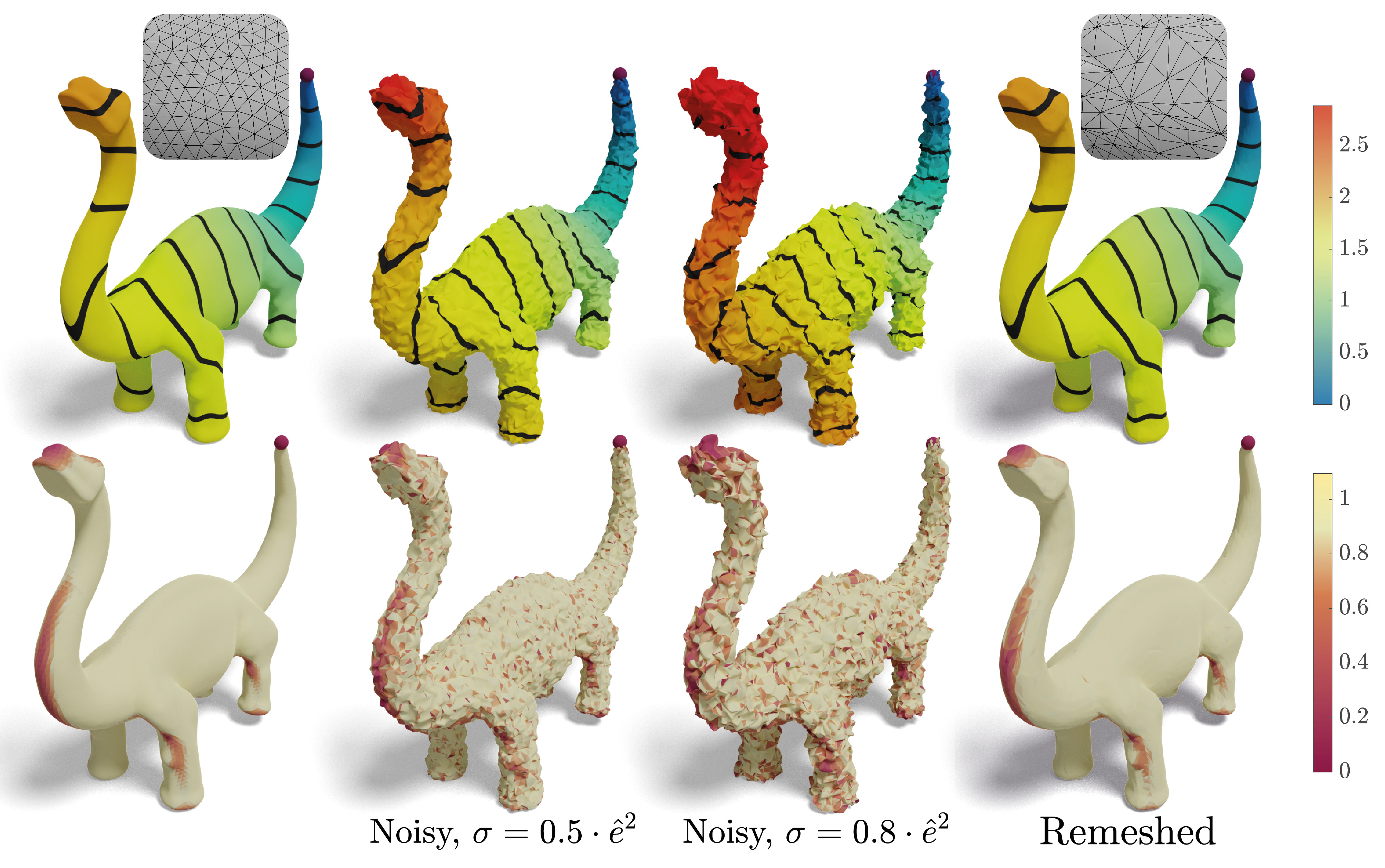}
    \caption{Robustness to noise and bad meshing, all distances computed with the same $\hat\smW$. Note the similarities of the distances and gradient norm, despite the large normal noise and badly shaped triangles.}
    \label{fig:robust} 
\end{figure}

\emph{Noise.}
Fig.~\ref{fig:robust} shows robustness to noise and bad meshing. We add Gaussian normal noise with $\sigma=0.5,0.8$ of the mean edge length, and use a remesh with highly anisotropic triangles and self-intersections. We show the distances and the gradient norm, all with the same $\alpha$. Note that the results are consistent between the different meshes. 

\begin{figure}
    \centering
    \includegraphics[width=\columnwidth]{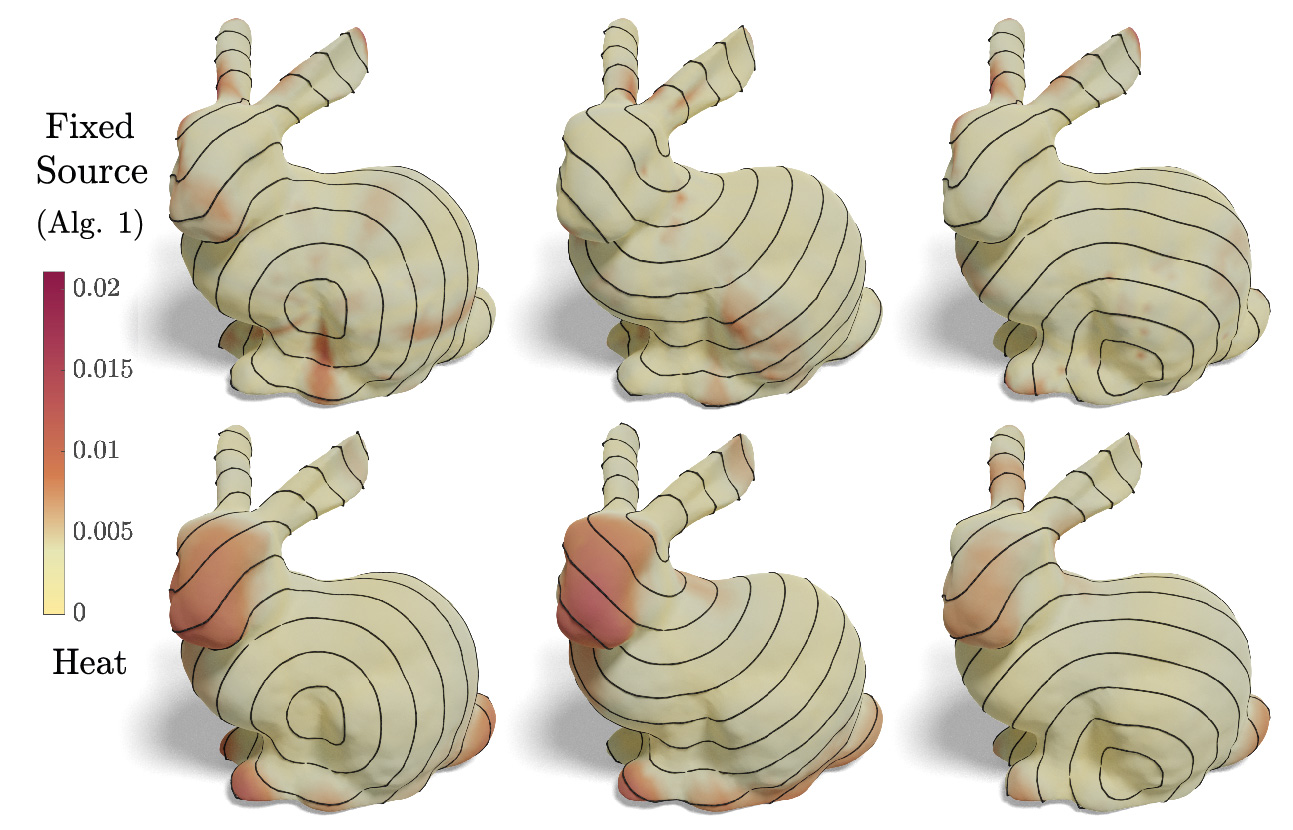}
    \caption{Violation of the symmetric property for $3$ source points. Note that while our method is not symmetric by construction, the symmetry error is lower than the symmetry error for the heat method.}
    \label{fig:symerr} 
\end{figure}

\emph{Symmetry error.}
Our Alg.~\ref{alg:admm} is not symmetric. Figure~\ref{fig:symerr} shows the symmetry error $\frac{1}{\sqrt{\mArea}}|d(x,y)-d(y,x)|$ for $3$ source points for our method and the heat method. Note that for all three points, the symmetry error is higher for the heat method.

\begin{figure}
    \centering
    \includegraphics[width=\columnwidth]{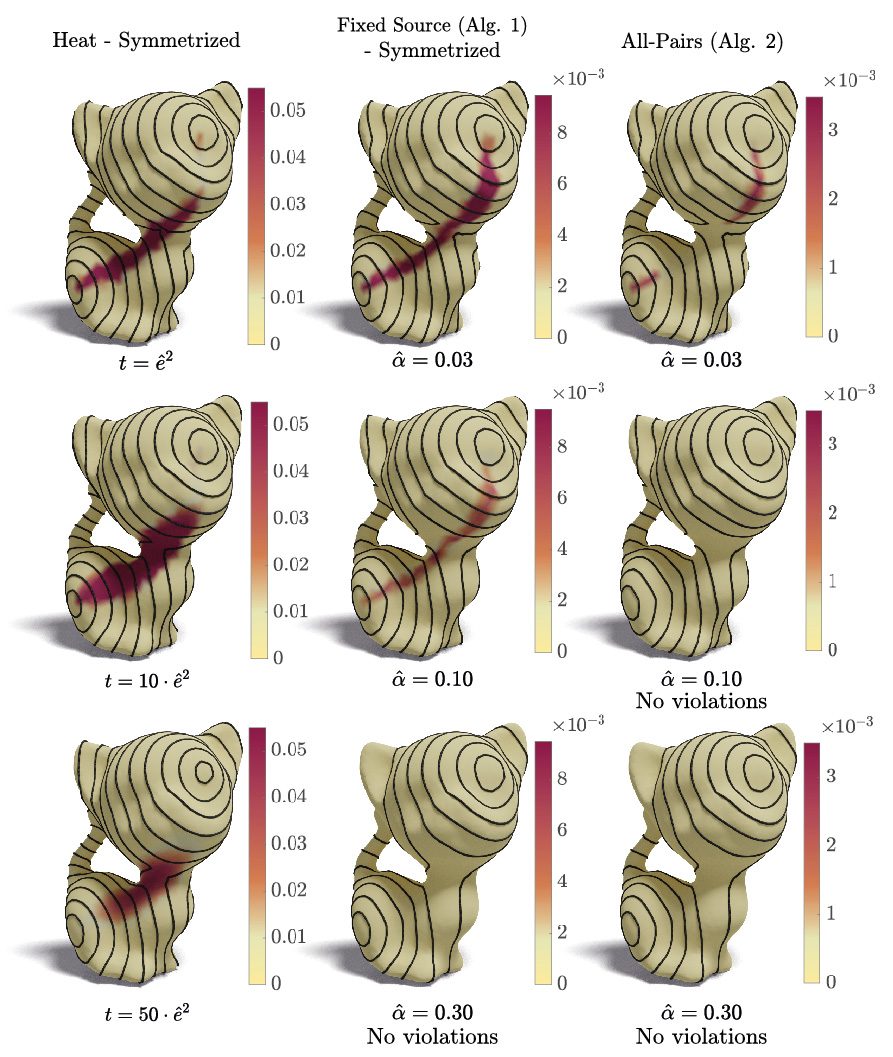}
    \caption{Triangle inequality violation. For a fixed pair of points (visualized with the distance isolines) we compute the triangle inequality error for all the other points. We compare the symmetrized heat method, our symmetrized method and the all-pairs formulation (which is symmetric by construction). We use a few smoothing weights for each approach. Note, that for our approach the violation reduces as the smoothing weight grows.}
    \label{fig:triangleinequality} 
\end{figure}

\emph{Triangle inequality error.}
Our method does not guarantee that triangle inequality holds, \ME{while EMD does}. However, experimentally it does hold for higher values of $\hat\smW$. Fig.~\ref{fig:triangleinequality} shows the triangle inequality error of a fixed pair of vertices with respect to every other vertex. We compare the heat method, Alg.~\ref{alg:admm}, and Alg.~\ref{alg:admmpm}. For the first two, we symmetrize the computed distance matrix. We show the results for three $t$ values for the heat method, and three values of $\hat\smW$ for our approach. We visualize the distance from the chosen point using isolines.
Note the difference in the error scaling between the two methods.
Further, note that for higher values of $\hat\smW$ our approach has no violations of the triangle inequality. Table~\ref{tab:trineqres} shows the percentage of triplets violating the triangle inequality for the same data. Note, that also when considering all the possible triplets, higher values of the smoothing parameter lead to less violations. 

\begin{table}
    \centering
    \smaller
    \caption{Percentage of triplets violating triangle inequality for the data in Fig.~\ref{fig:triangleinequality}. Note that for our approach higher values of $\smW$ lead to less violations.}
    \label{tab:trineqres}
    \small
        \begin{tabular}{  c |  c | c| c }             
             & Heat - Symmetrized & Fixed-Source - Symmetrized & All-Pairs \\
            \hline
             (a)  & 1.84 & 2.04 &  1.23 \\ 
             (b)  & 2.20 & 1.28 & 0.88 \\
             (c)  & 2.20 & 0.25 & 0.09\\
    \end{tabular}
    \vspace{-.2in}
\end{table}

\subsection{Volumetric Distances}
\begin{figure}
    \centering
    \includegraphics[width=\columnwidth]{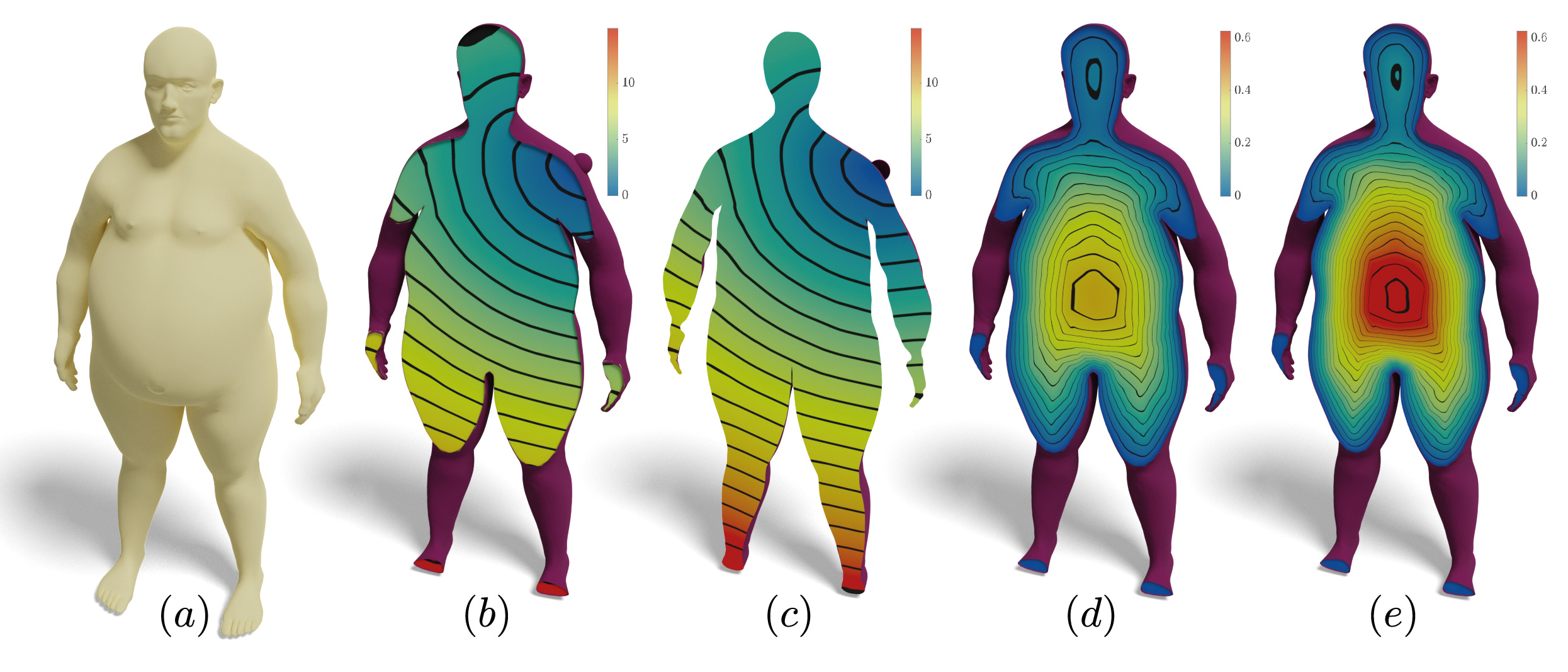}
    \caption{Dirichlet regularized volumetric distances. (a) The input tetrahedral mesh. (b,c) Two cuts showing the distance to a point on the shoulder. (d,e) Distance to the boundary, where (d) is more smoothed than (e), i.e. has a larger $\hat\smW$ value.}
    \label{fig:vol} 
\end{figure}

Our framework can compute distances on \emph{tetrahedral meshes}. We replace the standard mass matrix, gradient, and divergence operators with their volumetric versions, as implemented in gptoolbox~\cite{gptoolbox}. 
Figure \ref{fig:vol} demonstrates our Dirichlet regularized volumetric distance on a human shape (a). We show the distance from a point on the shoulder on two planar cuts (b,c), and the distance from the boundary using two $\hat\smW$ values (d,e).

\subsection{Example Application: Distance Function for Knitting}

\begin{figure}
    \centering
    \includegraphics[width=\columnwidth]{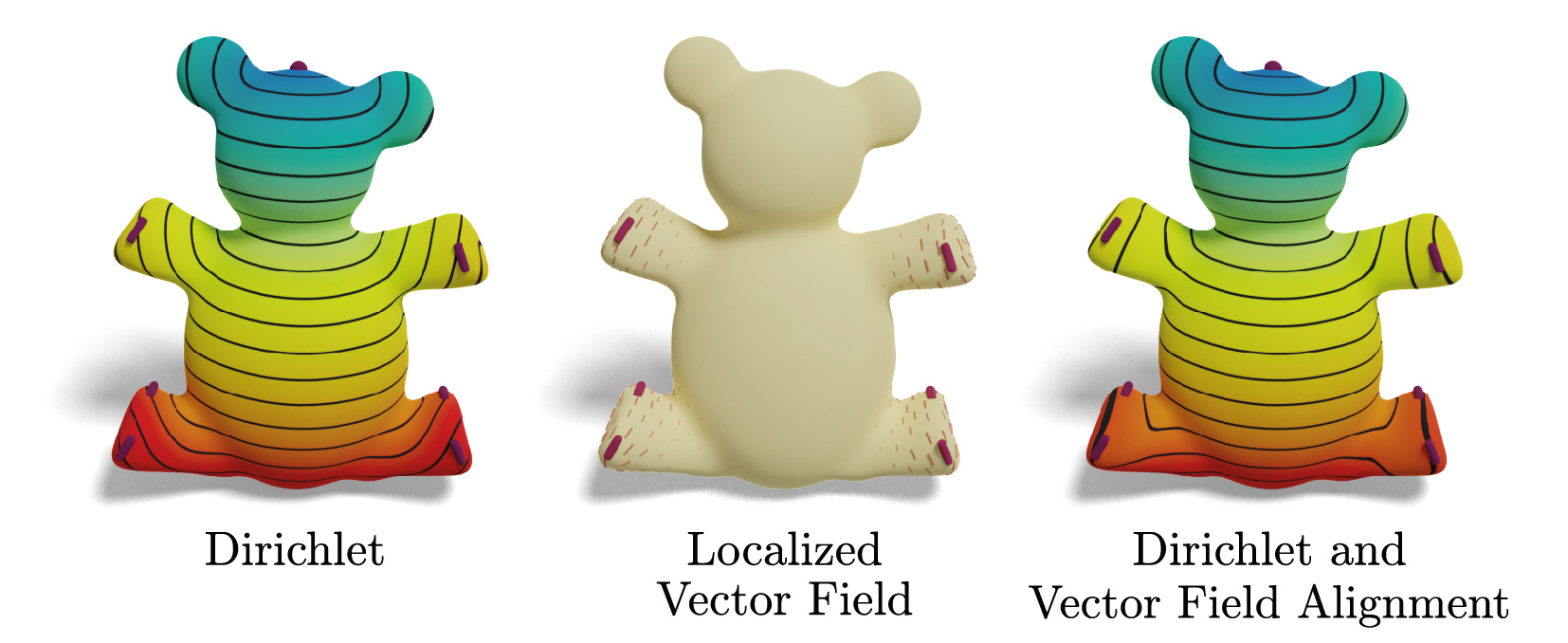}
    \caption{Vector field alignment of creases, useful for knitting applications.}
    \label{fig:knit}
\end{figure}

\ME{Some approaches for generating knitting instructions for $3$D models require a function whose isolines represent the knitting rows \cite{narayanan2018automatic, edelstein2022amigo}. Using the geodesic distance to an initial point (or a set of points) is a good choice since the stitch heights are constant, as are the distances between isolines. On the other hand, this choice limits the design freedom significantly, as designers and knitters have no control over the knitting direction on different areas of the shape. Using regularized distances with vector alignment solves this problem. For example, see Figure \ref{fig:vfa} and the C model. Using geodesic distances to the starting point will result in a non-symmetric shape (a) (see also \cite[Figure 10]{edelstein2022amigo}. By adding $2$ directional constraints (f), we obtain a function whose isolines respect the symmetries of the shape. Note that for the regularized distances, the gradient norm is no longer $1$ everywhere, and thus the distances between isolines is not constant. This can be addressed when knitting by using stitches of different heights. Figure \ref{fig:knit} shows how adding alignment to the teddy's arm and legs aligns the knitting rows with the creases. Crease alignment leads to better shaping~\cite[Section 9.3]{edelstein2022amigo}, and prevents over-smoothing of the knit model.}
 
\section{Conclusions and future work}
We presented a novel framework for constructing regularized geodesic distances on triangle meshes. We demonstrated the versatility of our approach by presenting three regularizers, analyzing them, and providing an efficient optimization scheme, as well as a symmetric formulation on the product manifold. The theoretical results and experiments in this work raise a number of interesting questions for future research. One of them is whether the functions $\pmdist_{\smW}(x,y)$ provide metrics in general, i.e., whether they satisfy the triangle inequality; we are not aware of results where geodesic distances can be regularized to have smooth metrics in $M\times M$.
Another theoretical question involves convergence of the minimizers in the Hessian energy-regularized problem, as discussed in Section \ref{sec:objectives}. 
\JS{Algorithmically, the ADMM algorithm from Section \ref{sec:admm_optim} easily generalizes to other convex functions $F_M$ (e.g., $L^1$ norms) in Equation \ref{eq:gen_reg_discrete}; recent theory on nonconvex ADMM also suggests that Algorithm~\ref{alg:admm} can be effective for nonconvex regularizers, possibly requiring large augmentation weights $\rho$ \cite{attouch2010proximal,hong2016convergence,wang2019global,zhang2019fundamental,zhang2019accelerating,ouyang2020anderson,gao2020admm,stein2022splitting}. 
}

\begin{acks}
Michal Edelstein acknowledges  funding  from  the  Jacobs Qualcomm Excellence Scholarship and the Zeff, Fine and Daniel Scholarship. 
Nestor Guillen was supported by the National Science Foundation through grant DMS-2144232.
The MIT Geometric Data Processing group acknowledges the generous support of Army Research Office grants W911NF2010168 and W911NF2110293, of Air Force Office of Scientific Research award FA9550-19-1-031, of National Science Foundation grants IIS-1838071 and CHS-1955697, from the CSAIL Systems that Learn program, from the MIT–IBM Watson AI Laboratory, from the Toyota–CSAIL Joint Research Center, from a gift from Adobe Systems, and from a Google Research Scholar award.
Mirela Ben-Chen acknowledges the support of the Israel Science Foundation (grant No. 1073/21), and the European Research Council (ERC starting grant no. 714776 OPREP). 
\METD{We use the repositories SHREC’07, SHREC’11, Windows 3D library, AIM@SHAPE, and Three D Scans, and thank Keenan Crane, Jan Knippers, Daniel Sonntag, and Yu Wang for providing additional models. 
We also thank Hsueh-Ti Derek Liu for his Blender Toolbox, used for the visualizations throughout the paper.} %
\end{acks}

\nocite{boyd2011distributed}
\nocite{boksebeld2022high}

\bibliographystyle{ACM-Reference-Format}
\bibliography{RGD}

\clearpage

\end{document}


\title[A Convex Optimization Framework for Regularized Geodesic Distances - Supplemental]{A Convex Optimization Framework for Regularized \\ Geodesic Distances - Supplemental Material}


\author{Michal Edelstein}
\orcid{0000-0001-9126-1617}
\affiliation{%
    \institution{Technion - Israel Institute of Technology}
    \city{Haifa}
    \country{Israel}}
\email{smichale@cs.technion.ac.il}

\author{Nestor Guillen} 
\orcid{0000-0002-4940-7595}
\affiliation{%
    \institution{Texas State University}
    \city{San Marcos}
    \state{TX}
    \country{USA}}
\email{nestor@txstate.edu}

\author{Justin Solomon} 
\orcid{0000-0002-7701-7586}
\affiliation{%
    \institution{Massachusetts Institute of Technology (MIT)}
    \city{Cambridge}
    \state{MA}
    \country{USA}}
\email{jsolomon@mit.edu}

\author{Mirela Ben-Chen} 
\orcid{0000-0002-1732-2327}
\affiliation{%
    \institution{Technion - Israel Institute of Technology}
    \city{Haifa}
    \country{Israel}}
\email{mirela@cs.technion.ac.il}

\maketitle

\section{Proofs regarding analytical solutions (Section 3.1)}

In this section we justify the formulas for the solutions to the problems in Section 3.1. Essential to these computations is the obstacle problem. For flat geometries or for general $M$ but with $\smW$ sufficiently small, the solution to problem 
(3) is the same as the solution to \emph{the obstacle problem with obstacle given by $d(x,E)$}. The obstacle problem in this case takes the form
\begin{align}\label{eqn:obstacle problem}
    \boxed{
    \begin{array}{rl}
    \textnormal{Minimize}_{\dist} & \smW \smE(\dist) -\int_\mM \dist(x)\;\text{dVol}(x) \\
    \textnormal{subject to} &  \dist(x) \leq d(x,E) \text{ for all } x\in M.
    \end{array}
    }   \tag{S1}
\end{align}
The equivalence between this problem and problem (3) is a classical fact in the case where $M$ is an open domain of $\mathbb{R}^n$ and $E = \partial M$, in this classical setting problem (3) is known as the elastic-plastic torsion problems. The equivalence in the situation $M$ is a Riemannian manifold is a more recent result and can be found in  [Générau et al. 2022].
In the general Riemannian case the equivalence between problem (3) and the obstacle problem  might not hold for all values of $\smW$, but it will hold for all $\smW$ smaller than some $\smW_0 = \smW_0(M)$.

\subsection{Analytical Solution for the Circle}  

We now prove the formula for the solution to the minimization problem in the case of $M = \mathbb{S}^1$ (Section 3.1). Recall that in Section 3.1 we have identified $\mathbb{S}^1$ with the real numbers modulo $2\pi$.

We claimed the solution where the source point is at $x=0$  is given by the formula
\begin{align}\label{eq:circle rgd example}
  \dist_\smW(x) = \left \{ \begin{array}{ll}
  \hat x & \text { if } 0 \leq \hat x \leq L\\
  \pi-\frac{1}{2}\smW-\frac{1}{2\smW}(\hat x-\pi)^2 & \text{ if } L\leq \hat x \leq 2\pi-L\\
  2\pi-\hat x & \text{ if } \hat x \geq L
  \end{array} \right. \tag{5}
\end{align}
(recall $\hat x$ is the unique representative of $x$ in the interval $[0,2\pi)$ modulo $2\pi$), where $L(\smW)$ is given by
\begin{align}\label{eq:circle example size smoothing region}
  L(\smW) =(\pi-\smW)_+. \tag{6}
\end{align}
We are going to show the function $u_\alpha$ is a solution to the obstacle problem \eqref{eqn:obstacle problem}, which in this case reduces to
\begin{align}\label{eq:obstacle problem circle}
    \boxed{
    \begin{array}{rl}
    \textnormal{Minimize}_{\dist} & \smW \int_0^{2\pi}|\dist'|^2\;dx -\int_0^{2\pi} \dist(x)\;dx \\
    \textnormal{subject to} &  \dist(x) \leq d(x,0) \text{ for all } x\in [0,2\pi]
    \end{array}
    }  \tag{S2}
\end{align}
The function $d(x,0)$ for $x\in [0,2\pi]$ is equal to
\begin{align*}
  d(x,0) = \min\{|x|,|x-2\pi|\} = \pi - |x-\pi|.
\end{align*}
The classical theory for the obstacle problem (see [Petrosyan et al.
2012])  says that if a function $\dist$ is of class $C^{1,1}$ in the entire domain (i.e. its gradient is Lipschitz continuous), is of class $C^2$ in the interior of $\{ u<d(x,0)\}$, and solves
\begin{align*}
  & \smW \dist'' + 1 \geq 0 \text{ in the sense of distributions},\\
  & \smW \dist'' + 1 = 0 \text{ in the interior of } \{ u < d(x,0) \}, 
\end{align*}
then that function $\dist$ will be the solution to the obstacle problem \eqref{eq:obstacle problem circle}. Let us verify this in our current example. First, by direct computation we can see $\dist$ has a continuous derivative in $(0,2\pi)$, and 
\begin{align}
  \dist'(x) = \left \{ \begin{array}{ll}
  1 & \text { if } 0 \leq \hat x \leq L\\
  -\frac{1}{\smW}(x-\pi) & \text{ if } L\leq \hat x \leq 2\pi-L\\
  -1 & \text{ if } \hat x \geq L
  \end{array} \right. \tag{S3}
\end{align}
We emphasize this function is continuous even at $x = L,2\pi-L$. Next, this function is twice differentiable away from $x= L,2\pi-L$ and in particular it is twice differentiable in the set $\{ u < d(x,0)\} = (L,2\pi-L)$. We have
\begin{align}
  \dist''(x) = \left \{ \begin{array}{ll}
  0 & \text { if } 0 \leq \hat x < L\\
  -\frac{1}{\smW} & \text{ if } L< \hat x < 2\pi-L\\
  0 & \text{ if } \hat x > L
  \end{array} \right. \tag{S4}
\end{align}
This shows that $\smW \dist''+1 = 0$ in $\{\dist<d(x,0)\}$. Lastly, since $\dist$ is differentiable everywhere this means that as a measure the function $\dist''$ is equal to $-(1/\smW)\chi_{(L,2\pi-L)}(x)$, $\chi$ denoting the indicator function. It follows that $\smW \dist''+1\geq 0$ in the sense of distributions. This proves that $\dist$ is indeed the solution to the obstacle problem, and in turn, of problem (3) in the case $M = \mathbb{S}^1$ and $E = \{0\}$.

That is, the function $\dist_{\smW}$ is of class $C^2$ away from $x= L,2\pi-L$ and $\dist_{\smW}''$ is continuous everywhere except at $L,2\pi-L$. From here follows that $\smW \dist''_{\smW} + 1$ is well defined as a measure, and that always $\geq 0$ and is exactly zero in the interval $(L,2\pi-L)$. This shows that $\dist_{\smW}$ solves
\begin{align*}
  \min\{ \smW \dist_{\smW}''+1,(\pi -|x-\pi|)-\dist_{\smW}\} = 0.
\end{align*}

Lastly, we prove the function $\dist_{\smW}(x,y)$ is indeed a metric. 

\subsection{Proof that $\dist_{\smW}(x,y)$ is a metric in $\mathbb{S}^1$}

By definition, we have $\dist_{\smW}(x,y) = \dist_{\smW}(x-y)$ where $\dist_\smW(x)$ is as in \eqref{eq:circle rgd example}. From here follows that $\dist_{\smW}(x,y)\geq 0$ for all $x,y$, since the function in \eqref{eq:circle rgd example} is non-negative. Moreover, the function in \eqref{eq:circle rgd example} only vanishes at $x$ equal to an integer multiple  of $2\pi$ (since in that case $\hat x = 0$, per the definition of $\hat x$), therefore $\dist_{\smW}(x,y) = 0$ only if $x-y$ is a multiple of $2\pi$, i.e. only if $x$ and $y$ correspond to the same point in $\mathbb{S}^1$.

To prove symmetry, simply observe that in \eqref{eq:circle rgd example} we have $\dist_{\smW}(x) = \dist_{\smW}(-x)$, so $\dist_{\smW}(x-y) = \dist_{\smW}(y-x)$. 

It remains to show $\dist_{\smW}(x,y)$ satisfies the triangle inequality. That is, we have to prove that for any $x,y,z$ we have
\begin{align*}
  \dist_{\smW}(x-y) \leq \dist_{\smW}(x-z) + \dist_{\smW}(z-y).
\end{align*}
By translation invariance (i.e. by symmetry) we may assume without loss of generality that $z=0$. Then, all we have to prove is that for all $x,y$ we have
\begin{align*}
  \dist_{\smW}(x-y) \leq \dist_{\smW}(x) + \dist_{\smW}(-y).
\end{align*}
Now, fix $y$ and consider the function
\begin{align*}
  v(x) := \dist_{\smW}(x-y)- \dist_{\smW}(-y).
\end{align*}
What we want to prove amounts to the inequality $v(x) \leq \dist_{\smW}(x)$. The function $v(x)$ satisfies the inequality
\begin{align*}
  v(x) \leq d(x,0),\;\text{ for all } x,
\end{align*}
as well as the differential inequality
\begin{align*}
  \smW v'' + 1 \geq 0.
\end{align*}
One well known characterization of the function $\dist_{\smW}(x)$ is that it is the largest function having these two properties. In this case we conclude that $\dist_{\smW}(x) \geq v(x)$, and the triangle inequality is proved. 

\subsection{Analytical Solution for the Disk}

To illustrate how our method handles other choices for the source set $E$, we take the flat 2D disk and consider the regularized distance to the boundary of the disk.

Using polar coordinates, we take $\sourceSet = \{ (r,\theta) | r=R\}$, and minimize
\begin{align*}
  \frac{\smW}{2}\int_{0}^{R}\int_{0}^{2\pi}|\mgrad \dist(r,\theta)|^2\;d\theta dr  - \int_{0}^{R}\int_{0}^{2\pi} \dist(r,\theta)r\;d\theta dr
\end{align*}
with the constraints
\begin{align*}
 & \dist(R,\theta)\leq 0 \text{ for all } \theta \in [0,2\pi], \text{ and } \\
 & |\nabla \dist(r,\theta)|\leq 1 \text{ for all } r\in[0,R),  \theta\in[0,2\pi].
\end{align*}
In this case, the solution is:

\begin{align}\label{eq:supplemental disk solution}
  \dist_{\smW}(x) = \left \{ \begin{array}{rl}
    -\frac{1}{4\smW}|x|^2 + R -\smW & \text{ if } |x|\leq 2\smW\\
    R-|x| & \text{ if }  2\smW < |x| \leq R
  \end{array}\right. \tag{S5}
\end{align}

\ME{To prove this formula, } we proceed similarly to the case of $\mathbb{S}^1$. 
This function is of class $C^{1,1}$, first note that its gradient is given by
\begin{align*}
  \nabla \dist_{\smW}(x) = \left \{ \begin{array}{rl}
    -\frac{1}{2\smW} x & \text{ if } |x|\leq 2\smW\\
    -\frac{x}{|x|} & \text{ if }  2\smW < |x| \leq R
  \end{array}\right.
\end{align*}
and this vector-valued function is continuous across $|x| = 2\smW$ (in fact, it is Lipscthiz continuous). Moreover, for the Laplacian of $\dist_{\smW}$ we have
\begin{align*}
  \Delta \dist_{\smW}(x) = \left \{ \begin{array}{rl}
    -\frac{1}{\smW} & \text{ if } |x|\leq 2\smW\\
    -\frac{1}{|x|} & \text{ if }  2\smW < |x| \leq R
  \end{array}\right.
\end{align*}
Accordingly, $\smW \Delta \dist_{\smW}+1\geq 0$ everywhere in the disk $\{ |x|<R\}$ and $\smW \Delta \dist_{\smW}+1 = 0$ exactly when $\dist_{\smW}<d(x,E) = R-|x|$. From here we conclude that the function $\dist_{\smW}$ given by \eqref{eq:supplemental disk solution} is the solution.

\begin{figure}
    \centering
    \includegraphics[width=\linewidth]{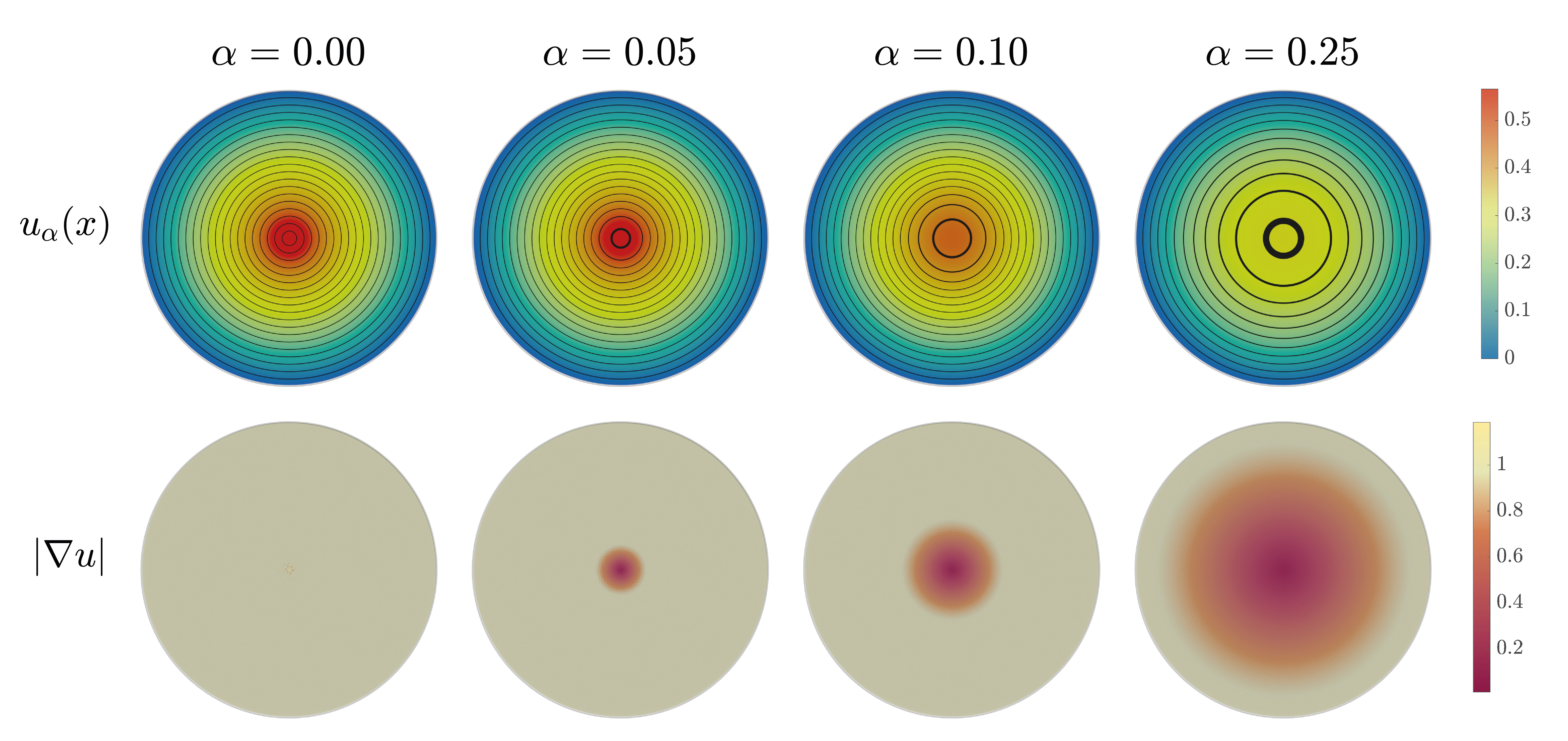}
    \caption{The analytical solution for the regularized geodesic distance using the Dirichlet regularizer on the disk (top). We also display the gradient norm, $| \mgrad \dist |$ (bottom). Note the different smoothing regions, whose width depends on $\smW$.}
    \label{fig:ana_disk}
\end{figure}

Figure~\ref{fig:ana_disk} shows the behavior of the function on the disk. Observe that as in the case of the circle, the solution has two regimes, one where it matches the distance function exactly, and one where it solves Poisson's equation $\Delta \dist = -1/\smW$. In this case, this results in the cone singularity being replaced by a concave quadratic function that is differentiable and only has a discontinuity in its second derivative. 

\section{Existence and uniqueness of the minimizer (Section 3)}

In our results, $M$ is a compact $C^\infty$ submanifold of $N$-dimensional Euclidean space $\mathbb{R}^N$, from where it inherits its Riemannian structure. The function $F(\xi,x)$, $F:\mathbb{R}^N\times \mathbb{R}^N \to \mathbb{R}$ is assumed of class $C^1$ in $(\xi,x)$. We make two further structural assumptions on $F$:  

\noindent 1) There are $p>1$ and $c_0,C_0$ positive such that
\begin{align*}
  c_0 |\xi|^p\leq F(\xi,x) \leq C_0|\xi|^p,\;\;\forall\;x,\xi\in\mathbb{R}^N
\end{align*}
2) The function $F$ is strictly convex in the first argument. This is meant in the following sense: given vectors $\xi_1 \neq \xi_2$ and $s\in (0,1)$ then we have the strict inequality for all $x$
\begin{align*}
  F((1-s)\xi_1+s\xi_2,x) < (1-s)F(\xi_1,x)  + sF(\xi_2,x).
\end{align*}
From these assumptions follows in particular that $F(\xi,x) \geq 0$ for all $\xi$ and $x$, and $F(\xi,x) = 0$ only if $\xi=0$. Observe that these assumptions include all $F$'s of the forms
\begin{align*}
  F(\xi,x) = |A(x)\xi|^p
\end{align*}
where $p>1$ and $A(x)$ is a smooth positive definite matrix whose eigenvalues are uniformly bounded away from zero and infinity.  

Now we prove the existence and uniqueness for the general minimization problem. The problem (see problem (3)) is a constrained minimization problem in the Sobolev space $W^{1,p}(M)$.
\begin{align}
\label{eq:General Minimization Problem}
  \boxed{
    \begin{array}{rl}
  \textnormal{Minimize}_u & \smW \int_{\mM}F(\nabla \dist,x)\;\text{dVol}(x)-\int_{\mM}\dist\;\text{dVol}(x) \\
    \textnormal{subject to} &  u \in W^{1,p}(M) \\ & |\nabla \dist(x)|\leq 1 \text{ for all } x\in M\setminus \sourceSet \\
	& \dist(x) \leq 0  \text{ for all } x \in \sourceSet.
    \end{array}} \tag{3}
\end{align}
The space $W^{1,p}(M)$ ($1\leq p <\infty$) is defined as follows
\begin{align*}
  W^{1,p}(M) := \left \{ u:M \to \mathbb{R} \;\mid\; \begin{array}{l} \nabla u \text{ exists as a distribution} \\ \int_{M}|u|^p+|\nabla u|^p\;dx<\infty \\ \end{array} \right \}
\end{align*}

Here, $E\subset M$ is a non-empty closed subset of $M$. For us, the case of chief interest is when $\sourceSet = \{ \source\}$ for a given $\source \in M$.

In what follows, we will denote the objective functional by $J_\smW$,
\begin{align*}
  J_\smW(\dist) := \smW \int_{\mM}F(\nabla \dist,x)\;\text{dVol}(x)-\int_{\mM}\dist\;\text{dVol}(x)
\end{align*}

We now prove the existence and uniqueness first theorem stated in Section 3. 
\begin{manualtheorem}{3.1}
  There is a unique minimizer for problem (3).
\end{manualtheorem}

\begin{proof}
  Consider a minimizing sequence $\{\dist_k\}_k$. First, we claim that without loss of generality, we can assume that for each $k$,
  \begin{align}\label{eqn:non-nonpositivity}
    \max_\mM \dist_k \geq 0.\tag{S7}
  \end{align}
  Indeed, if for some $k_0$ we had $\dist_{k_0}$ is non-positive in all of $\mM$, it would follow that
  \begin{align*}
    J_\smW(\dist_{k_0}) \geq 0 = J_\smW(0). 
  \end{align*}
  Thus the minimizing sequence will remain a minimizing sequence if we replace every non-positive element of the sequence with the zero function. 
  
  Henceforth, we assume our sequence $\dist_k$ is such that \eqref{eqn:non-nonpositivity} holds for all $k$. In this case, as the $\dist_k$ are all $1$-Lipschitz, it follows that
  \begin{align*}
    \dist_k(x) \geq \max_\mM \dist_k(x) - \text{diam}(\mM) \geq -\text{diam}(\mM) \text{ for all } k.
  \end{align*}
  A similar argument---using that $\max_{x\in \sourceSet} \dist_k(x)\leq 0$ for all $k$---provides the inequality in the opposite direction. In conclusion,
  \begin{align*}
    \|\dist_k\|_{L^\infty(\mM)} \leq \text{diam}(\mM)\;\text{ for all } k. 
  \end{align*}
  It follows that the sequence $\{\dist_k\}_k$ is $1$-Lipschitz and uniformly bounded. Then, by the Arzela-Ascoli theorem there is a subsequence $\dist_k'$ and a function $\dist_*$ in $\mM$ such that
  \begin{align*}
    \|\dist_k'-\dist_*\|_{L^\infty(\mM)} \to 0 \text{ as } k\to\infty. 
  \end{align*}
  Moreover, without loss of generality (we can always pass to another subsequence where this holds)) we also have
  \begin{align*}
    \nabla \dist_k' \to \nabla \dist_* \text{ in weak-}L^p(M).
  \end{align*}
  In particular, 
  \begin{align*}
    \lim\limits_{k} \int_{M} \dist_k'(x)\;\text{dVol}(x) = \int_{M} \dist_*(x)\;\text{dVol}(x).
  \end{align*}
In this case, due to the weak convergence of $\nabla \dist_k'$, as well as the convexity of $F$ in the first argument and its two-sided pointwise bounds, we conclude that 
  \begin{align*}
    \liminf\limits_{k}\int_{M}F(\nabla \dist_k',x)\text{dVol}(x) \geq \int_{M}F(\nabla \dist_*,x)\text{dVol}(x)
  \end{align*}
  Putting everything together, we ahve shown that
  \begin{align*}
      \liminf\limits_k J_\smW(\dist_k') \geq J_\smW(\dist_*).
  \end{align*}
  Since the $1$-Lipschitz constraint as well as the constraint $\dist_k(x)\leq 0$ for all $x\in \sourceSet$ are preserved by the uniform convergence, it follows that $\dist_*$ is admissible. Moreover, the last liminf inequality says $\dist_*$ achieves the minimum of $J_\smW$ among all admissible functions, so $\dist_*$ is a minimizer for the problem.

  The uniqueness follows from the strict convexity through a standard argument, which we review for completeness: suppose there are two separate minimizers $\dist_0$ and $\dist_1$. For each $s \in [0,1]$ let
  \begin{align*}
    \dist_s = (1-s)\dist_0 + s \dist_1.
  \end{align*}
  From the convexity assumption on $F$ we know that
  \begin{align*}
    F(\nabla \dist_s,x) \leq (1-s)F(\nabla \dist_0,x) + s F(\nabla \dist_1,x).
  \end{align*}
  In terms of the functional, this gives us
  \begin{align*}
    J_\smW(\dist_s) \leq (1-s)J_\smW(\dist_0) + J_\smW(\dist_1) 
  \end{align*}
  by merely integrating the pointwise inequality, and at the same time
  \begin{align*}
    J_\smW(\dist_s) \geq (1-s)J_\smW(\dist_0) + J_\smW(\dist_1) 
  \end{align*}
  Since $\dist_0$ and $\dist_1$ are minimizers and $\dist_s$ is admissible for every $s\in [0,1]$. This can only happen if 
  \begin{align*}
    F(\nabla \dist_s,x) = (1-s)F(\nabla \dist_0,x) + s F(\nabla \dist_1,x) \;\forall\;x \in M,
  \end{align*}
  and by the assumption, this can only happen if $\nabla \dist_0 = \nabla \dist_1$ at every $x$. As $\dist_0$ and $\dist_1$ have to agree at least at one point, $\source$, this means that $\dist_0=\dist_1$.
\end{proof}

\section{Convergence to the geodesic distance (Section 3)}

In this section we prove the convergence theorem from Section 3.
\begin{manualtheorem}{3.2}    
  The functions $\dist_\alpha$ converge uniformly to $d(\cdot,\sourceSet)$,
  \begin{align*}
      \|\dist_\smW-d(\cdot,\sourceSet)\|_{L^\infty} \to 0 \text{ as } \smW \to 0^+.
  \end{align*}
        
\end{manualtheorem}

\begin{proof}
We make use of an elementary but often used fact in nonlinear PDE that says that compactness plus uniqueness of the limiting points of a sequence in turn guarantees convergence of the whole sequence.\footnote{if the sequence failed to converge as whole, there would be some $\delta>0$ an infinite subsequence such that $\dist_{\smW_k'}$ stays a distance at least $\delta$ from $d(x,\source)$, leading to a contradiction} Concretely, and in two parts, we are going to show 1) that given any sequence $\smW_k \to 0^+$ we can pass to a subsequence $\smW_k'$ which converges uniformly to some function $\dist_*$
\begin{align*}
  \|\dist_{\smW_k'}-\dist_*\|_{L^\infty(\mM)} \to 0 \text{ as } k \to \infty,
\end{align*}
and subsequently that 2) whatever function $\dist_*$ is obtained as one of these limits will have to be a minimizer for problem (1). Since that problem has as its unique solution, then $u_* = d(x,\source)$ for all such subsequences.

Indeed, first, note that the $1$-Lipschitz constraint and the fact that $\dist_{\smW_k}(\source) = 0$ for all $k$ implies that the sequence $\dist_{\smW_k}$ lies in a compact subset of $C(M)$. Therefore, there is a subsequence $\smW_k'$ and a $1$-Lipschitz function $\dist_* \in C(M)$ such that 
\begin{align*}
  \dist_{\smW_k'} \to \dist_* \text{ uniformly in } M.
\end{align*}
Now, let $\phi:M\to \mathbb{R}$ be a $1$-Lipschitz function such that $\phi(x_0) \leq 0$. Since $\phi$ is admissible for \eqref{eq:General Minimization Problem} for every $\smW_k'$, it follows that
\begin{align*} 
  -\int_{M} \dist_{\smW_k'}\text{dVol}(x) & \leq  \smW_k' \int_M F(\nabla \dist_{\smW_k'},x)\text{dVol}(x)-\int_{M}\dist_{\smW_k'}\;dx \\
  & \leq \smW_k' \int_M F(\nabla \phi,x)\text{dVol}(x)-\int_{M}\phi\;dx.
\end{align*}
On the other hand, by the $1$-Lipschitz constraint
\begin{align*}
  \int_M F(\nabla \phi,x)\text{dVol}(x) \leq C\text{Vol}(M).
\end{align*}
This means in particular that
\begin{align*}
  -\int_{M} \dist_*\text{dVol(x)} & = \lim\limits_{k} \int_{M}\dist_{\smW_k'}\text{dVol}(x)\\
  & \leq \lim \limits_{k}\left \{ \smW_k' \int_M F(\nabla \phi,x)\text{dVol}(x)-\int_{M}\phi\;dx \right \}\\
  & = -\int_{M}\phi(x)\;dx.
\end{align*}
This shows that $\dist_*$ solves the minimization problem (1), and this problem has a known solution, so $\dist_* = d(\cdot,\sourceSet)$. In summary we have shown that given any sequence $\smW_k\to 0$ there is a subsequence $\smW_k'$ such that $\dist_{\smW_k'} \to d(\cdot,\sourceSet)$ uniformly in $M$, finishing the proof.  
\end{proof}

\section{Analytical solution for the Hessian regularizer in $1$D (Hessian Regularizer, Section 4.2)}

In $1D$, the Hessian energy is the same as the bilaplacian energy, and the optimization problem is:
\begin{equation*}
\label{eq:gdH}
    \begin{array}{rl}
    \textnormal{Minimize}_{\dist} & \frac{\smW}{2} \int_0^{2\pi}|\dist''(x)|^2\;\text{d}x -\int_0^{2\pi} \dist(x)\;\text{d}x \\
    \textnormal{subject to} &  |\dist'(x)|\leq 1 \text{ for all } x\in (0,2\pi) \\
	& \dist(0) \leq 0. \\
    \end{array}
\end{equation*}

The minimizer $\dist(x)$ is (for $x\in [0,2\pi]$):
\begin{align*}
  \dist(x) = \left \{ \begin{array}{ll}
  x & \text { if } 0 \leq x \leq \pi- c\\
  \begin{array}{ll}
   \frac{1}{24\alpha}(x-\pi)^4 - \frac{c^2}{4\alpha}(x-\pi)^2 \\
   +\pi - c + \frac{5c^4}{24\alpha} \\
  \end{array} 
  & \text{ if } \pi- c \leq x \leq \pi+ c\\
  2\pi-x & \text{ if } x \geq \pi+ c
  \end{array} \right.
\end{align*}
where $c=\sqrt[3]{3\alpha}$.

\ME{Note that the function and smoothing region are different than the ones in the Dirichlet regularizer case.}

\section{Existence and uniqueness of a minimizer (product manifold formulation, Section 6)} 

In Section 6 we introduced the following problem.
\begin{align}\label{eq:product manifold minimization problem}
  \boxed{
  \begin{array}{rl}
  \textnormal{Minimize} & \smW \mathcal{E}_{M\times M}(\pmdist)- \int_{M\times M} \pmdist(x,y)\;\text{dVol}(x,y)\\
  \textnormal{subject to} & \pmdist \in W^{1,2}(M\times M)\\
  &|\nabla_1 \pmdist(x,y)| \leq 1 \text{ in }  \{(x,y)\;\mid\; x\neq y\}\\
  & |\nabla_2 \pmdist(x,y)| \leq 1 \text{ in }  \{(x,y)\;\mid\; x\neq y\}\\
  & \pmdist(x,y) \leq 0 \text{ on } \{ (x,y) \;\mid\; x=y\}
  \end{array}
  } \tag{12}
\end{align}

\begin{manualtheorem}{6.1}
  There is a unique minimizer in problem \eqref{eq:product manifold minimization problem}.
\end{manualtheorem}

\begin{proof}
  At the big picture level this proof is basically the same as that of existence and uniqueness of a minimizer for problem \eqref{eq:General Minimization Problem}. We only highlight the points where things are different.

  Therefore, take a minimizing sequence $\pmdist_k$. Arguing similarly as before we can assume without loss of generality that 
  \begin{align*}
    \max \limits_{M\times M}\pmdist_k \geq 0 \text{ for all } k.
  \end{align*} 
  Now, $\pmdist_k$ is $1$-Lipschitz in each variable $x$ and $y$, separately, so, if for some $k$ $(x_0,y_0)$ is a point where $\pmdist_k(x_0,y_0) \geq 0$, then for all other $(x,y)$ we have
  \begin{align*}
    \pmdist_k(x,y) & \geq \pmdist_k(x,y_0) - d(y,y_0)\\
      & \geq \pmdist_k(x_0,y_0) - d(x,x_0)-d(y,y_0) \\
      & \geq \pmdist_k(x_0,y_0) - 2\text{diam}(M).
  \end{align*}
  On the other hand, since $\pmdist_k(x,x) \leq 0$ for all $x$ and $y$, we have, using the $1$-Lipschtz condition in the first variable
  \begin{align*}
    \pmdist_k(x,y) & \leq \pmdist_k(x,x) + d(x,y) \\
      & \leq d(x,y) \leq \text{diam}(M).
  \end{align*}
  Putting all this together we have
  \begin{align*}
    \|\pmdist_k\|_{L^\infty(M)} \leq 2\text{diam}(M) \text{ for all } k.
  \end{align*}   
  This means our sequence $\{\pmdist_k\}_k$ is uniformly bounded and equicontinuous (in fact, uniformly Lipschitz) in the compact space $M\times M$. 
  
  By the Arzela-Ascoli theorem, there is a subsequence $\pmdist_k'$ and a function $\pmdist_*$ in $M \times M$ such that $\pmdist_k$ converges uniformly to $\pmdist_*$. In particular, this function $\pmdist_k$ will be Lipschitz and the inequalities
  \begin{align*}
    |\nabla_1 \pmdist_*(x,y)| \leq 1 \text{ and } |\nabla_2 \pmdist_*(x,y)|
  \end{align*}
  hold for a.e. $(x,y) \in M\times M$. Moreover, $\pmdist_*(x,x) \leq 0$ for all $x\in M$. This shows that $\pmdist_*$ is admissible for problem \eqref{eq:product manifold minimization problem}. At the same time, the uniform convergence of the $\pmdist_k$ and the compactness of $M\times M$ imply that
  \begin{align*}
    \lim\limits_k \int_{M\times M} \pmdist_k'(x,y)\text{dVol}(x,y) = \int_{M\times M} \pmdist_*(x,y)\text{dVol}(x,y).
  \end{align*}
  Lastly, passing to another subsequence $\pmdist_k''$ if needed, we have
  \begin{align*}
    & \liminf\limits_k \int_{M\times M}|\nabla_1\pmdist_k''|^2+|\nabla_2\pmdist_k''|^2\text{dVol}(x,y)\\
    & \geq \int_{M\times M}|\nabla_1\pmdist_*|^2+|\nabla_2\pmdist_*|^2\text{dVol}(x,y).
  \end{align*}
  From here it follows that $\pmdist_*$ is a minimizer for problem \eqref{eq:product manifold minimization problem}.

  Uniqueness is proved again making use of the convexity of the functional. 
  
\end{proof}

A consequence of the uniqueness theorem is the symmetry of the minimizers $\pmdist_\smW$:
\begin{manualtheorem}{6.2}\label{lem:product formulation symmetry}
  The function $\pmdist_\smW(x,y)$ is symmetric in $x$ and $y$.
\end{manualtheorem}

\begin{proof}
  This is a direct consequence of the uniqueness of the minimizer to problem (12) as well as the symmetry of the under the transformation $(x,y) \mapsto (y,x)$. Indeed, given $\smW$ define the function
  \begin{align*}
    \mathscr{v}_\smW(x,y) := \pmdist_\alpha(y,x),
  \end{align*}
  Then it is clear that $\mathscr{v}$ is still admissible for problem (12) and 
  \begin{align*}
    & \smW \smE_{M\times M}(\pmdist_\smW) -\int_{\mM \times \mM} \pmdist_\smW(x,y)\;\text{dVol}(x,y) \\
    & = \smW \smE_{M\times M}(\mathscr{v}) -\int_{\mM \times \mM} \mathscr{v}(x,y)\;\text{dVol}(x,y),
  \end{align*}
  so that $\mathscr{v}$ also achieves the minimum of problem (12). Since there is only one minimizer, $\mathscr{v} = \pmdist_{\smW}$ and the lemma follows.
\end{proof}

\section{Convergence to the full geodesic distance (product manifold formulation, Section 6)} 

In this section we will make use of the following characterization of the geodesic distance function $d(x,y)$. 

\begin{align}\label{eq:characterization of the full geodesic distance}
  \boxed{\begin{array}{rl}
  \textnormal{Minimize} & - \int_{M\times M} v(x,y)\;\text{dVol}(x,y)\\
  \textnormal{subject to} & |\nabla_1 v(x,y)| \leq 1 \text{ in }  \{(x,y)\;\mid\; x\neq y\}\\
  & |\nabla_2 v(x,y)| \leq 1 \text{ in }  \{(x,y)\;\mid\; x\neq y\}\\
  & v(x,y) \leq 0 \text{ on } \{ (x,y) \;\mid\; x=y\}
  \end{array}} \tag{S8}
\end{align}
The problem \eqref{eq:characterization of the full geodesic distance} clearly resembles problem \eqref{eq:product manifold minimization problem}. Accordingly, the proof Theorem 6.3 (just as the proof of Theorem 3.2, Supplemental 3) will consist in using compactness and show all limit points of the sequence $\pmdist_{\smW}$ as $\smW \to 0$ have to be just $d(x,y)$.
\begin{theorem}
  As $\smW\to 0$, we have
  \begin{align*}
    \|d(x,y)-\pmdist_{\smW}(x,y)\|_{L^\infty(M\times M)} \to 0.
  \end{align*}
\end{theorem}

\begin{proof}
  Let $\smW_k \to 0$ be any sequence. The sequence $\{\pmdist_{\smW_k}\}_k$ is uniformly Lipschitz, accordingly, there is a subsequence $\{\pmdist_{\smW_k}'\}_k$ and a function $\pmdist_*(x,y)$ such that $\pmdist_{\smW_k}' \to \pmdist_*$ uniformly in $M\times M$ as $k\to \infty$. We are going to show $\pmdist_*$ must be the geodesic distance.

  Indeed, let $\Phi(x,y) \to \mathbb{R}$ be any smooth admissible function for problem (12). Then, for any $\smW>0$ we have
  \begin{align*}
     & - \int_{M\times M} \pmdist_{\smW_k}'(x,y)\;\text{dVol}(x,y)\\ & \leq  \smW_k'\mathcal{E}_{M\times M}(\pmdist_{\smW_k}') - \int_{M\times M} \pmdist_{\smW_k}'(x,y)\;\text{dVol}(x,y) \\
     & \leq \smW_k's\mathcal{E}_{M\times M}(\Phi) - \int_{M\times M} \Phi(x,y)\;\text{dVol}(x,y).
  \end{align*}
Taking the limit $\smW_k' \to 0$ with the last inequality, it follows that 
  \begin{align*}
     & - \int_{M\times M} \pmdist_*(x,y)\;\text{dVol}(x,y)\\ & \leq  - \int_{M\times M} \Phi(x,y)\;\text{dVol}(x,y).
  \end{align*}
  Since $\Phi$ is an arbitrary admissible function, it follows that $\pmdist_* = d(x,y)$. This proves that $\pmdist_k$ converges uniformly to $d(x,y)$, and in turn, by the same argument as in the proof of Theorem 3.2 (Supplemental 3) that $\pmdist_{\smW} \to d(x,y)$ as $\smW\to 0$.
\end{proof}

\section{Non-quadratic regularizers}

The functional $F(\xi,x)$ used for the regularizer term in the minimization problem (3) allows for quite general norms or powers of norms. Using a non-isotropic norm from the ambient space, one obtains $F$'s that have no dependence on $x$ but manifest behavior that is sensitive to the position and orientation of $M$. We illustrate this with some numerical experiments with
\begin{align}\label{eq:non quadratic F}
  F(\xi,x) = \|\xi\|_\infty^2,  \tag{S9}
\end{align}
which satisfies all of the assumptions for Theorems 3.1 and 3.2 (as discussed at the start of Supplemental 2). Although \eqref{eq:non quadratic F} involves a square, it is different from a quadratic polynomial on the entries of $\xi$. In fact, $F$ in \eqref{eq:non quadratic F} is not differentiable for all $\xi$, this can be seen by writing $F$ in terms of the components of the vector $\xi$, if $\xi^t = (\xi_1,\xi_2,\xi_3)$ then 
\begin{align*}
  \|\xi\|_\infty = \max \{\xi_1^2,\xi_2^2,\xi_3^2\}.
\end{align*}
This is a convex function of $(\xi_1,\xi_2,\xi_3)$, it is smooth in the open set $\{ |\xi_i| \neq |\xi_j| \text{ if } i \neq j\}$, but it is not differentiable along the boundary of this set.

\begin{figure}
    \centering
    \includegraphics[width=\linewidth]{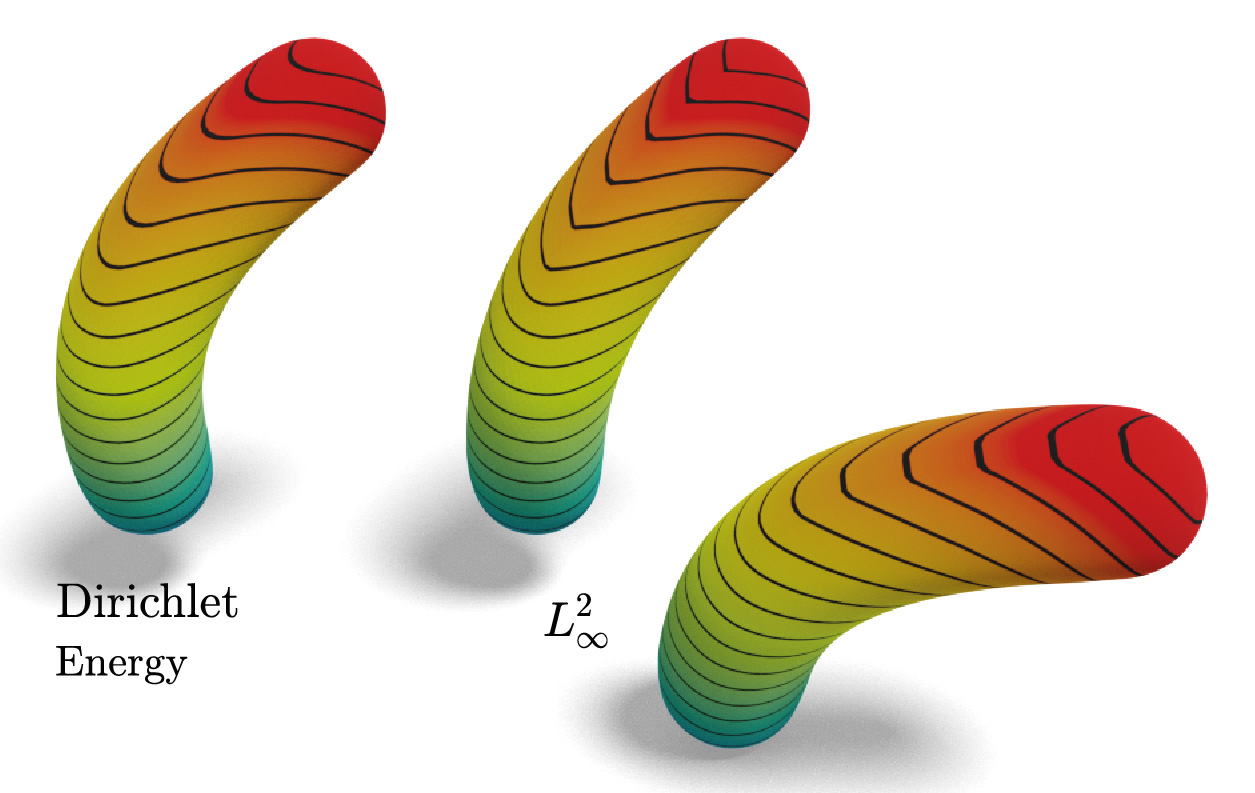}
    \caption{Non-quadratic regularizer example. We compare the results between the quadratic Dirichlet energy (left) to using the squared $L_\infty$ norm on two meshes with different orientations (right). See the text for details.}
    \label{fig:linf}
\end{figure}

In Fig.~\ref{fig:linf}, we see how the regularized geodesics with $\mathcal{E}$ using \eqref{eq:non quadratic F} look like for different orientations. Observe the anisotropic effects as the pipe is rotated as well as the flatter level curves.  

\section{Distances to a fixed source ADMM derivation (Section 5.4)} 

In Section 5.4 we present the augmented Lagrangian used to derive the ADMM algorithm.
 
\begin{align*}
    \begin{array}{rl}
    L(\dist,\admmlm,\admmz) &=   -\mva^T \dist + \frac{\smW}{2} \dist^T \optW \dist + \displaystyle \sum_{f \in \mF} \chi(|\admmz_f| \le 1) + \\
    & \displaystyle \sum_{f \in \mF} a_f \admmlm_f^T ((\mgradd \dist)_f - \admmz_f) + \frac{\admmpen \sqrt{\mArea}}{2} \displaystyle \sum_{f \in \mF} a_f |(\mgradd \dist)_f - \admmz_f|^2,
    \end{array}
\end{align*}
where $a_f$ is the area of the face $f$, $\admmpen \in \mathbb{R}$ is the penalty parameter, and $\admmlm \in \mathbb{R}^{3\nF}$ is the dual variable or Lagrange multiplier.

The ADMM algorithm iterates between three stages~[Boyd et al.
2011, Section 3]: 
$\dist$-minimization, $\admmz$-minimization, and updating the dual variable. Where using this formulation, both $\dist$ and $\admmz$ have closed-form solutions. 

The ADMM algorithm alternates between these three steps:
\begin{enumerate}
    \item $\dist^{k+1} = [\smW \optW+\admmpen \sqrt{\mArea} \mW]^{-1} [\mva - \mdivd \admmlm^k + \admmpen \sqrt{\mArea} \mdivd \admmz^k]$ 

    \item $\admmz_f^{k+1} = \text{Proj} (\frac{1}{\admmpen \sqrt{\mArea} } \admmlm_f^k + (\mgradd \dist^{k+1})_f, \mathbb{B}^3)$
        \text{ for all } $f\in \mF $\\
    \item $\admmlm^{k+1} = \admmlm^k + \admmpen \sqrt{\mArea} (\mgradd \dist^{k+1} - \admmz^{k+1})$,
\end{enumerate}
where $\text{Proj}(z_f\!\in\!\xR^3,\mathbb{B}^3)$ is equal to $z_f/|z_f|$ if $|z_f|>1$, and $z_f$ otherwise. 

We consider our algorithms to have converged when $\|r^{k}\| \le \epsilon^{pri}$ and $\|s^{k}\| \le \epsilon^{dual}$, where $r^{k}$ and $s^{k}$ are the primal and dual residuals, resp. And $\epsilon^{pri}, \epsilon^{dual}$ are the primal and dual feasibility tolerances, resp. These quantities can be computed as follows:
\begin{equation*}
\begin{array}{rl}
    r^{k} &= \sqrt{\mfaM} \mgradd \dist^{k} - \sqrt{\mfaM} \admmz^{k} \\
    s^{k} &= \rho \mdivd (\admmz^{k}- \admmz^{k-1}) \\
    \epsilon^{pri} &= \sqrt{3 \nF} \epsilon^{abs} \mArea + \epsilon^{rel} \sqrt{\mArea} \text{max}{(\|\sqrt{\mfaM} \mgradd \dist^{k}\|,\|\sqrt{\mfaM} \admmz^{k}\|)} \\
    \epsilon^{dual}&= \sqrt{\nV} \epsilon^{abs} \mArea + \epsilon^{rel} \sqrt{\mArea} \|\mdivd \admmlm\|.
    \end{array}
\end{equation*}
In all our experiments, we set $\epsilon^{abs} = 5\cdot 10^{-6}, \epsilon^{rel} = 10^{-2}$, and  $\admmpen = 2 $. We define $\admmpen$, the residuals and feasibility tolerances such that they are scale-invariance, as explained in Section 7.1.

In addition, to accelerate the convergence, we also use the varying penalty parameter and over-relaxation, exactly as described in [Boyd
et al. 2011, Sections 3.4.1, 3.4.3].

\section{Symmetric All-Pairs ADMM Derivation (Section 6.2)} 

Our discrete optimization problem, as introduced in Section 6.2, is:
\begin{align*}
\label{eq:dgdpm}
    \begin{array}{rl}
    \textnormal{Minimize}_{\pmdist} & 
     - \mva^T \pmdist \mva \,+ \\
     & \frac{1}{2}\smW \, \text{Tr}\left(\mvaM \big(\pmdist^T \mW \pmdist + 
       \pmdist \mW \pmdist^T \big) \right) \\
    \textnormal{subject to} 
    & |(\mgrad \pmdist_{(i,\cdot)} )_f |\leq 1 \quad \text{ for all } f\in \mF, i\in \mV\\
    & |(\mgrad \pmdist_{(\cdot,j)} )_f |\leq 1 \quad \text{ for all } f\in \mF, j\in \mV\\
	& \pmdist_{i,i} \leq 0 \,\,\, \qquad \qquad \text{ for all } i\in \mV,
    \end{array}
\end{align*}
where $X_{i,j}$ denotes the $(i,j)$-th element of a matrix $X$,  $X_{(i,\cdot)}$ denotes the $i$-th row, and $X_{(\cdot,j)}$ the $j$-th column. 

Our derivation is based on the consensus problem  [Boyd et al .
2011, Section 7], where we split $\pmdist$ into two variables $\admmx, \admmr \in \mathbb{R}^{\nV \times \nV}$ to represent the gradient along the columns and rows, and use a consensus auxiliary variable $\admmzd  \in \mathbb{R}^{\nV \times \nV}$ to ensure consistency. We also add two
auxiliary variables $\admmZ, \admmzq \in \mathbb{R}^{3\nF\times \nV}$ representing the gradients along the columns and rows, i.e., $\mgradd \admmx, \mgradd \admmr$. We enforce the diagonal constraint on the consensus variable $\admmzd$ to avoid solving huge linear systems. This leads to the following optimization problem:

\begin{equation*}
\label{eq:dgdpm}
    \begin{array}{rl}
    \textnormal{Minimize}_{\pmdist} & 
    -\frac{1}{2} \mva^T \admmx \mva - \frac{1}{2} \mva^T \admmr \mva \,+ \\
     & \frac{1}{2}\smW \, \text{Tr}\left(\mvaM \big(\admmx^T \mW \admmx + 
       \admmr^T \mW \admmr \big) \right) \,+\\
      & \displaystyle \sum_{f \in \mF} \sum_{i \in \mV} \chi(|(\admmZ_{(\cdot,i)})_f| \le 1) \,+ \\
      & \displaystyle \sum_{f \in \mF} \sum_{i \in \mV} \chi(|(\admmzq_{(\cdot,i)})_f| \le 1) \\
    \textnormal{subject to} 
    & (\mgradd \admmx_{(\cdot,i)})_f =(\admmZ_{(\cdot,i)})_f \quad \text{ for all } f\in \mF, i\in \mV\\
    & (\mgradd \admmr_{(\cdot,i)})_f =(\admmzq_{(\cdot,i)})_f \quad \text{ for all } f\in \mF, i\in \mV\\
    & \admmx = \admmzd \\
    & \admmx = \admmzd \\
    & \admmr = \admmzd^T \\
	& \pmdist_{i,i} \leq 0 \,\,\,\, \qquad \qquad \qquad \text{ for all } i\in \mV \\
    & \admmzd \ge 0 ,
    \end{array}
\end{equation*}

where $\chi(\admmZ_{(\cdot,i)})_f| \le 1) = \infty$ if $\admmZ_{(\cdot,i)})_f| > 1$ and $0$ otherwise.

The corresponding augmented Lagrangian is:

\begin{align*}
    \begin{array}{rl}
    L(\pmdist,\admmlmY,\admmZ) &=   
    -\frac{1}{2} \mva^T \admmx \mva - \frac{1}{2} \mva^T \admmr \mva \,+ \\
     & \frac{1}{2}\smW \, \text{Tr}\left(\mvaM \big(\admmx^T \mW \admmx + 
       \admmr^T \mW \admmr \big) \right) \,+\\
      & \displaystyle \sum_{f \in \mF} \sum_{i \in \mV} \chi(|(\admmZ_{(\cdot,i)})_f| \le 1) \,+ \\
      & \displaystyle \sum_{f \in \mF} \sum_{i \in \mV} \chi(|(\admmzq_{(\cdot,i)})_f| \le 1) \,+ \\
      & \text{Tr}\left(\mvaM \big(\admmlmY^T \mfaM (\mgradd \admmx - \admmZ) + \admmlmu^T \mfaM (\mgradd \admmr - \admmzq) \big) \right) \,+ \\
      & \frac{\admmpen_1 \sqrt{\mArea}}{2} \text{Tr}\left(\mvaM    \big(\mgradd \admmx - \admmZ \big)^T \mfaM    \big(\mgradd \admmx - \admmZ \big) \right) \,+ \\
      & \frac{\admmpen_1 \sqrt{\mArea}}{2} \text{Tr}\left(\mvaM   \big(\mgradd \admmr - \admmzq \big)^T \mfaM    \big(\mgradd \admmr - \admmzq \big) \right) \,+ \\
      & \text{Tr}\left(\admmlmm^T \big(\admmx - \admmzd\big) \mvaM \right) +
       \text{Tr}\left (\admmlmn^T \big(\admmr - \admmzd^T\big) \mvaM \right) \,+ \\
      & \text{Tr}\left( \admmlmm^T \big(\admmx - \admmzd\big) \mvaM \right) \,+ \\
      & \frac{\admmpen_2 \sqrt{\mArea^{-1}}}{2} \text{Tr}\left( \big( \admmx - \admmzd \big)^T \mvaM   \big(\admmx - \admmzd \big) \mvaM \right) \,+ \\
      & \frac{\admmpen_2 \sqrt{\mArea^{-1}}}{2} \text{Tr}\left( \big( \admmr - \admmzd^T \big)^T \mvaM   \big(\admmr - \admmzd^T \big) \mvaM \right), 
    \end{array}
\end{align*}
where $\admmpen_1, \admmpen_2 \in \mathbb{R}$ are the penalty parameters, and $\admmlmY, \admmlmu \in \mathbb{R}^{3\nF \times \nV}$, $\admmlmm, \admmlmn \in \mathbb{R}^{\nV \times \nV}$ are the dual variables.

The ADMM algorithm for this optimization problem consists of three stages. In the first stage, we optimize for $\admmZ, \admmr$. In the second step, we minimize the auxiliary variables $\admmZ, \admmzq, \admmzd$. Finally, in the third step, we update the dual variables added in the augmented Lagrangian.

\begin{enumerate}
    \item 
    \begin{align*}
        \begin{array}{rl}
         \admmx^{k+1} &=   \left( (\smW + \admmpen_1 \sqrt{\mArea})\mW + \admmpen_2 \sqrt{\mArea^{-1}} \mvaM \right) ^{-1} \,\\
           & \big( \frac{1}{2} \mva \mva^T \mvaM^{-1} - \mdivd \admmlmY^k \,+\\
           & \admmpen_1 \sqrt{\mArea} \sqrt{\mArea} \mdivd \admmz^k  - \mvaM \admmlmm^k + \admmpen_2 \sqrt{\mArea^{-1}} \mvaM \admmzd^k \big) \\
        \end{array}
    \end{align*}

    \begin{align*}
        \begin{array}{rl}
        \admmr^{k+1} &=
        \left( (\smW + \admmpen_1 \sqrt{\mArea})\mW + \admmpen_2 \sqrt{\mArea^{-1}} \mvaM \right) ^{-1}  \,\\
         & \big( \frac{1}{2} \mva \mva^T \mvaM^{-1} - \mdivd \admmlmu^k \,+\\
         & \admmpen_1 \sqrt{\mArea} \mdivd \admmzq^k  - \mvaM \admmlmn^k + \admmpen_2 \sqrt{\mArea^{-1}} \mvaM \admmzd^{kT} \big) 
         \end{array}
    \end{align*}

    \item $(\admmZ_{(\cdot,i)}^{k+1})_f =
        \\ \quad \text{Proj} \left( \frac{1}{\admmpen_1\sqrt{\mArea} } (\admmlmY_{(\cdot,i)}^k)_f + (\mgradd \admmx_{(\cdot,i)}^{k+1})_f,\mathbb{B}^3 \right)  \text{ for all } i \!\in \!\mV, f\!\in\!\mF $\\

        $(\admmzq_{(\cdot,i)}^{k+1})_f =
        \\ \quad \text{Proj} \left( \frac{1}{\admmpen_1 \sqrt{\mArea}} (\admmlmu_{(\cdot,i)}^k)_f + (\mgradd 
        \admmr_{(\cdot,i)}^{k+1})_f,\mathbb{B}^3 \right) \text{ for all } i\! \in\! \mV, f\!\in\! \mF $\\

        $\admmzd^{k+1} = \text{max}\left(\frac{\admmlmm^k+\admmlmn^{kT}}{2\admmpen_2 \sqrt{\mArea^{-1}}} + \frac{\admmx^{k+1}+\admmr^{kT}}{2}, 0\right)$ \\
        $\admmzd^{k+1}_{i,i} = 0$ \text{ for all } $i \in \mV$\\
        
    \item $\admmlmY^{k+1} = \admmlmY^k + \admmpen_1 \sqrt{\mArea} \big( \mgradd \admmx^{k+1} - \admmz^{k+1} \big)$\\
        $\admmlmu^{k+1} = \admmlmu^k + \admmpen_1 \sqrt{\mArea} \big( \mgradd \admmr^{k+1} - \admmzq^{k+1} \big)$ \\
        $\admmlmm^{k+1} = \admmlmm^k + \admmpen_2 \sqrt{\mArea^{-1}} \big( \admmx^{k+1} - \admmzd^{k+1} \big)$ \\
        $\admmlmn^{k+1} = \admmlmn^k + \admmpen_2 \sqrt{\mArea^{-1}} \big( \admmr^{k+1} - \admmzd^{kT} \big)$ \\

\end{enumerate}

Similarly to Section 5.4, the first steps include solving a linear system with the same coefficient matrix, which can be pre-factored to accelerate the computation. 

We consider our algorithms to have converged when $\|r^{k}\| \le \epsilon^{pri}$ and $\|s^{k}\| \le \epsilon^{dual}$, where $r^{k}$ and $s^{k}$ are the primal and dual residuals, resp. And $\epsilon^{pri}, \epsilon^{dual}$ are the primal and dual feasibility tolerances, resp. These quantities can be computed as follows:

\begin{equation*}
\begin{array}{rl}
    r^{k} &= \sqrt{\mfaM} \mgradd \dist^{k} - \sqrt{\mfaM} \admmz^{k} \\
    s^{k} &= \rho \mdivd (\admmz^{k}- \admmz^{k-1}) \\
    \epsilon^{pri} &= \sqrt{3 \nF} \epsilon^{abs} \mArea + \epsilon^{rel} \text{max}{(\|\sqrt{\mfaM} \mgradd \dist^{k}\|,\|\sqrt{\mfaM} \admmz^{k}\|)} \\
    \epsilon^{dual} &= \sqrt{\nV} \epsilon^{abs} \mArea^2 + \epsilon^{rel} \|\sqrt{\mfaM} \mdivd \admmlmY\| 
    \end{array}
\end{equation*}
and equivalently for $\admmr, \admmzq, \admmlmu$. The residuals for the consensus part are as follows:
\begin{equation*}
\begin{array}{rl}
    r_1^{k} &= \mvaM (\admmx^{k} - \admmzd^{k}) \mvaM\\
    r_2^{k} &= \mvaM (\admmr^{k} - \admmzd^{Tk}) \mvaM\\
    s^{k} &= \rho_2 \mvaM (\admmzd^{k}- \admmzd^{k-1}) \mvaM \\
    \epsilon_1^{pri} &= \sqrt{\nV} \epsilon^{abs} + \epsilon^{rel} \text{max}{(\|\mvaM \admmx^{k} \mvaM\|, \|\sqrt{\mfaM} \admmz^{k} \sqrt{\mvaM}\mvaM  \admmzd^{k} \mvaM\|)} \\ 
    \epsilon_2^{pri} &= \sqrt{\nV} \epsilon^{abs} \sqrt{\mArea^3} + \epsilon^{rel} \text{max}{(\|\mvaM \admmr^{k} \mvaM\|, \|\mvaM \admmzd^{Tk} \mvaM\|)} \\
    \epsilon^{dual} &= \sqrt{\nV} \epsilon^{abs} \mArea + \frac{\epsilon^{rel}}{2} (\|\sqrt{\mvaM} \admmlmm \sqrt{\mvaM} \|+\|\sqrt{\mvaM} \admmlmn \sqrt{\mvaM} \|). 
    \end{array}
\end{equation*}
We set $\epsilon^{abs} = 10^{-6}, \epsilon^{rel} = 2 \cdot 10^{-4}$ and $\admmpen_1 = \admmpen_2 = 2 $ in all our experiments. Note that both the penalty variables, the residuals and feasibility thresholds are defined to be  scale-invariance, as explained in section 7.1.

In addition, to accelerate the convergence, we also use the varying penalty parameter and over-relaxation, exactly as described in [Boyd
et al. 2011, Sections 3.4.1, 3.4.3]. 

\section{Additional Results}
\subsection{Additional Examples}
\begin{figure*}
    \centering
    \includegraphics[width=2\columnwidth]{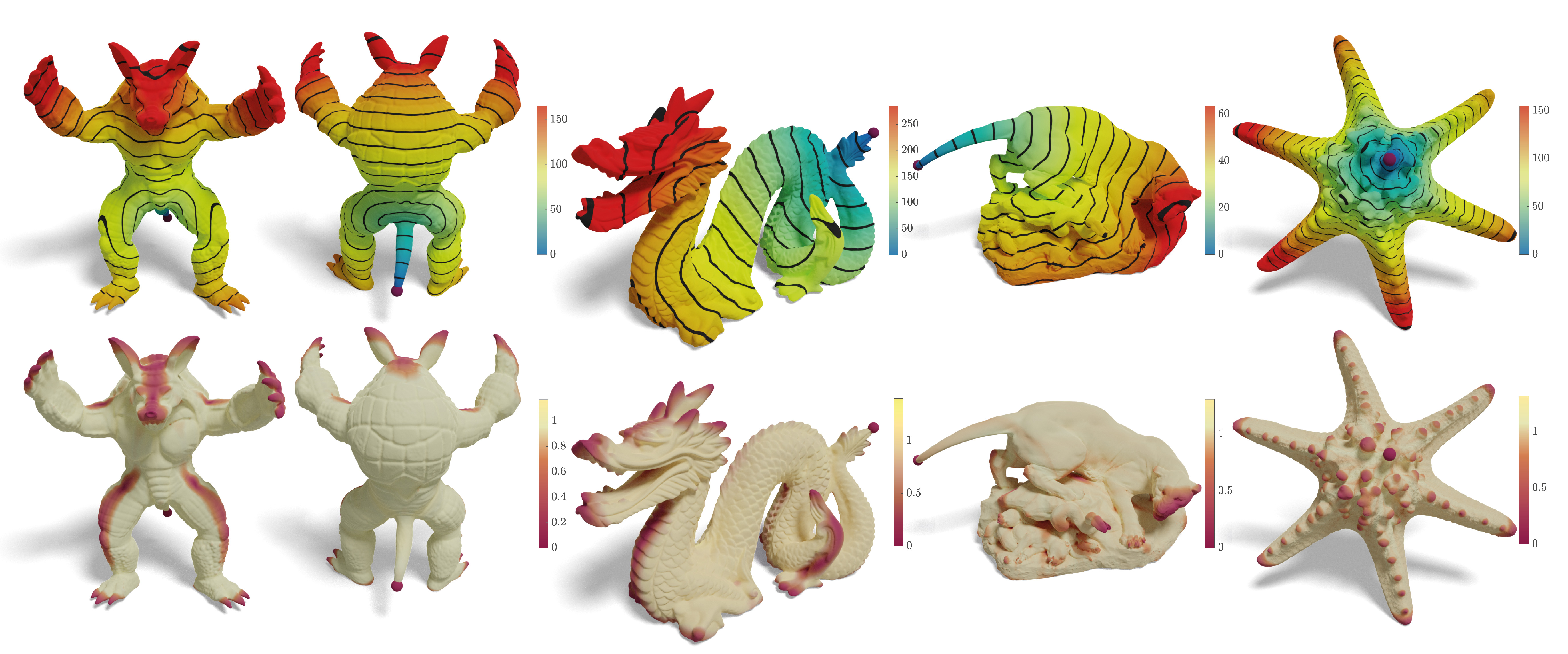}
    \caption{The distance isolines and gradient norm with Dirichlet regularization for various meshes.}
    \label{fig:more}
\end{figure*}

Figure \ref{fig:more} shows more examples of our fixed source (Alg. 1) method for the meshes in Table 1, Section 5.4.

\subsection{Representation Error in a Spectral Reduced Basis}

\begin{figure}
    \centering
    \includegraphics[width=\columnwidth]{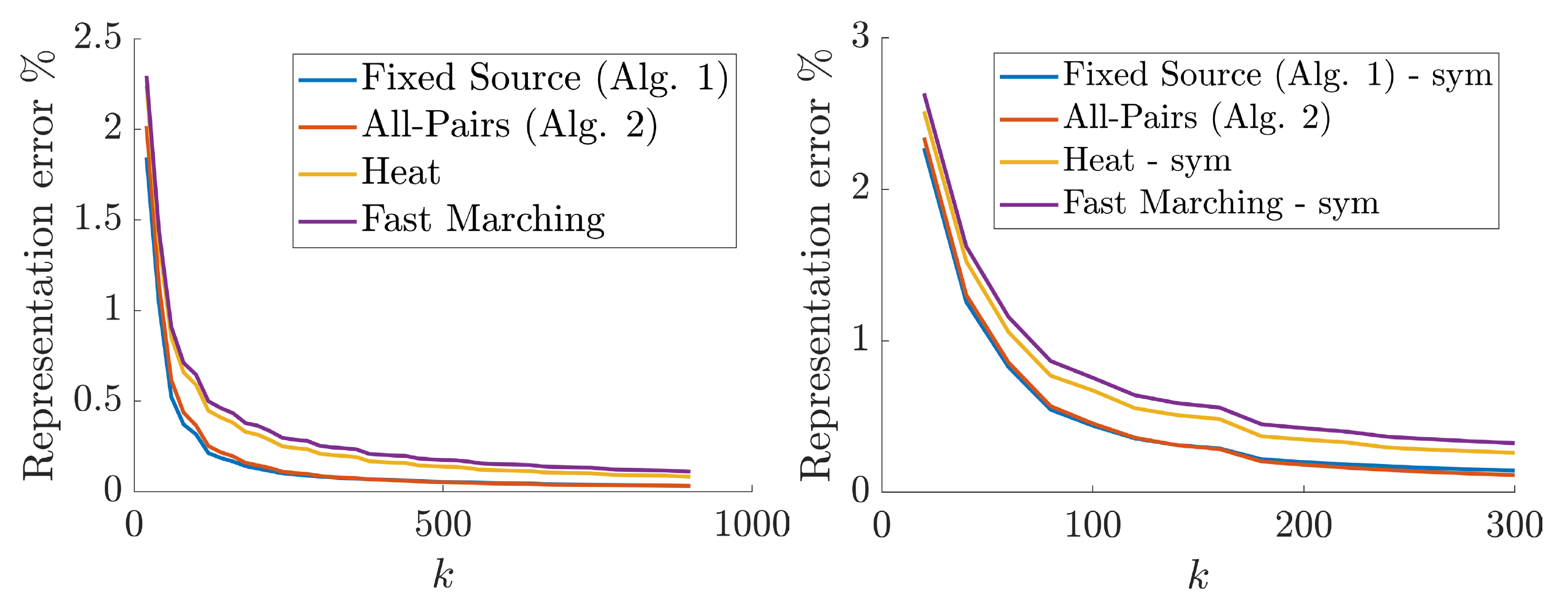}
    \caption{Comparison of the representation error of the Dirichlet regularized distances in a reduced spectral basis. See the text for details.}
    \label{fig:spectral}
\end{figure}

Smoother functions are better represented in a reduced basis comprised of the eigenvectors of the Laplace-Beltrami operator. Namely, they require less basis functions for the same representation error. In Fig.~\ref{fig:spectral} (left) We compare the representation error in a reduced basis of our approach, the heat method, and fast marching. Note that our approach, both the fixed source (Alg. 1) and the all-pairs (Alg 2.) formulations, achieves the lowest error (indicating that the functions are smoothest in this sense). Similarly, we compare the symmetric formulations by symmetrizing our fixed source method, the heat method and the Fast Marching results, see Fig.~\ref{fig:spectral} (right). Here we project on the eigenvectors of the LB operator on the \emph{product manifold}. Here as well we achieve a lower error than the alternatives.
The experiment was done on the ``pipe'' mesh, where we computed the full distance matrix between all pairs of vertices. For Fig~\ref{fig:spectral} (left) we projected each column of the distance matrix (i.e., the distance from a single source vertex), and computed the mean of the representation errors. For Fig~\ref{fig:spectral} (right), we projected the full distance matrix on the eigenvectors of the LB operator on the product manifold.

\subsection{Additional Results on Various Triangulations}

\begin{figure}[b]
    \centering
    \includegraphics[width=0.935\columnwidth]{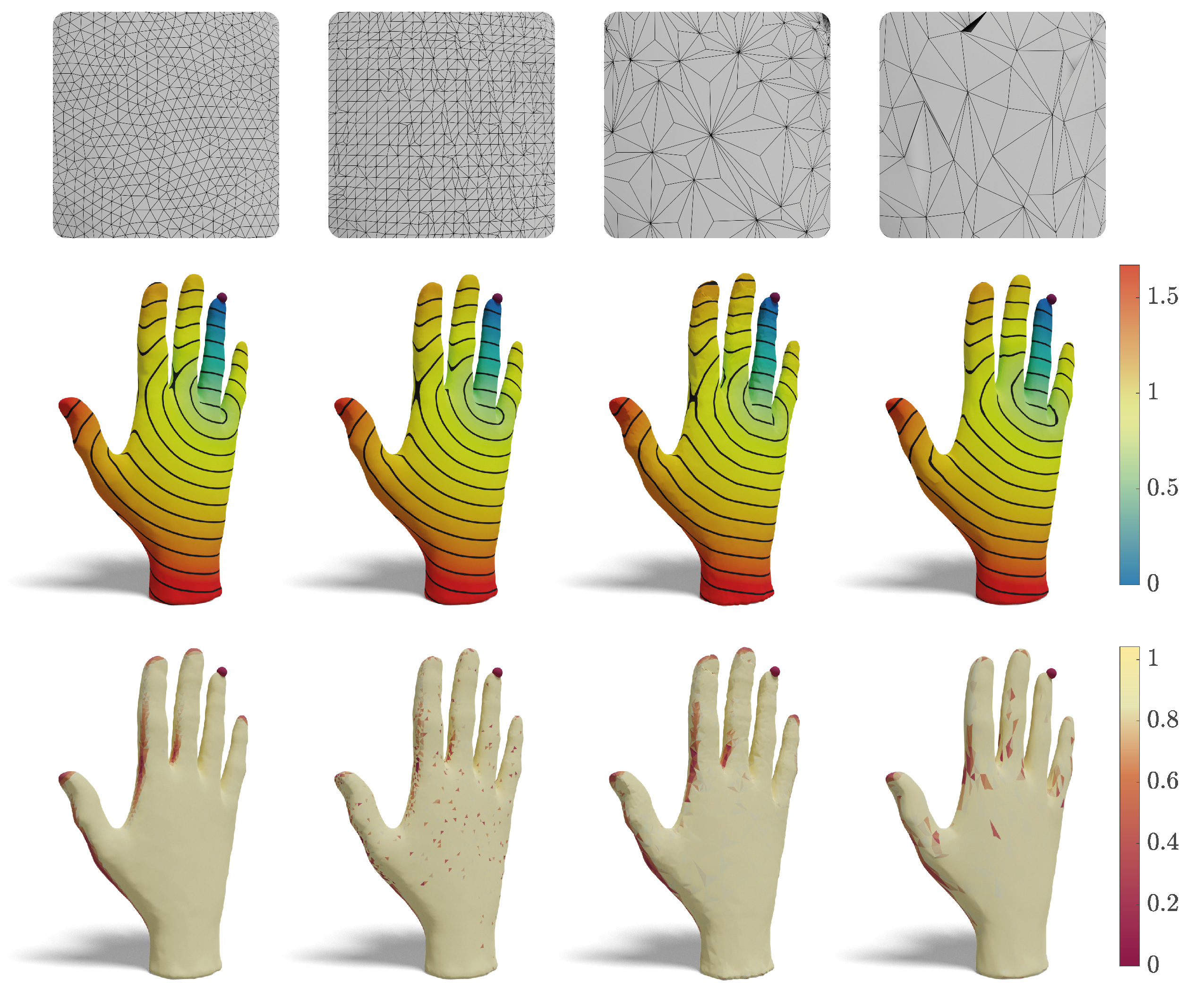}
    \caption{The regularized geodesic distance using the Dirichlet regularizer for various triangulations. For each triangulation, we display the connectivity (top), the isoline of the distance (middle) and the gradient norm, $| \mgrad \dist |$ (bottom). Note that the results are qualitatively similar for all the triangulations.}
    \label{fig:triangulations}
\end{figure}

\ME{To further demonstrate the robustness of our algorithm, we show additional results on low-quality triangulations in Figure \ref{fig:triangulations}. The leftmost column corresponds to a uniform triangulation and the other three to non-uniform triangulations. Note that the results remain similar for the different triangulations.}

\subsection{Timings for the All-Pairs Formulation}
\ME{Table~\ref{tab:trineqrestimings} shows the running times for computing the all-pairs distances on the cat model. We compare the heat method, computed using Geometry Central [Sharp et al. 2019] (using the precomputation speed-up), our fixed source formulation (Alg. 1) and our all-pairs approach (Alg. 2). Note that Alg. 2 has a higher memory overhead than Alg. 1, because we are working with large dense matrices. Therefore, in our non-optimized Matlab implementation we may run out of memory for large meshes. We believe that a more careful implementation can improve this considerably.}

\begin{table}
    \centering
    \caption{Timings in seconds for the All-Pairs distance computation on the cat model, $| \mF | = 3898$, Figure 10 (main paper).}
    \label{tab:trineqrestimings}
    \small
        \begin{tabular}{  c |  c | c| c }             
             & Heat - Symmetrized & Fixed-Source - Symmetrized & All-Pairs \\
             & [sec] & [sec] & [sec] \\
            \hline
             (a)  & 0.77 & 101.625 &  1124.312 \\ 
             (b)  & 0.77 & 59.583 &  837.063 \\
             (c)  & 0.76 & 37.4745 &  794.9549 \\
    \end{tabular}
\end{table}

\subsection{Quadratic Finite Elements}
\begin{figure}
    \centering
    \includegraphics[width=\linewidth]{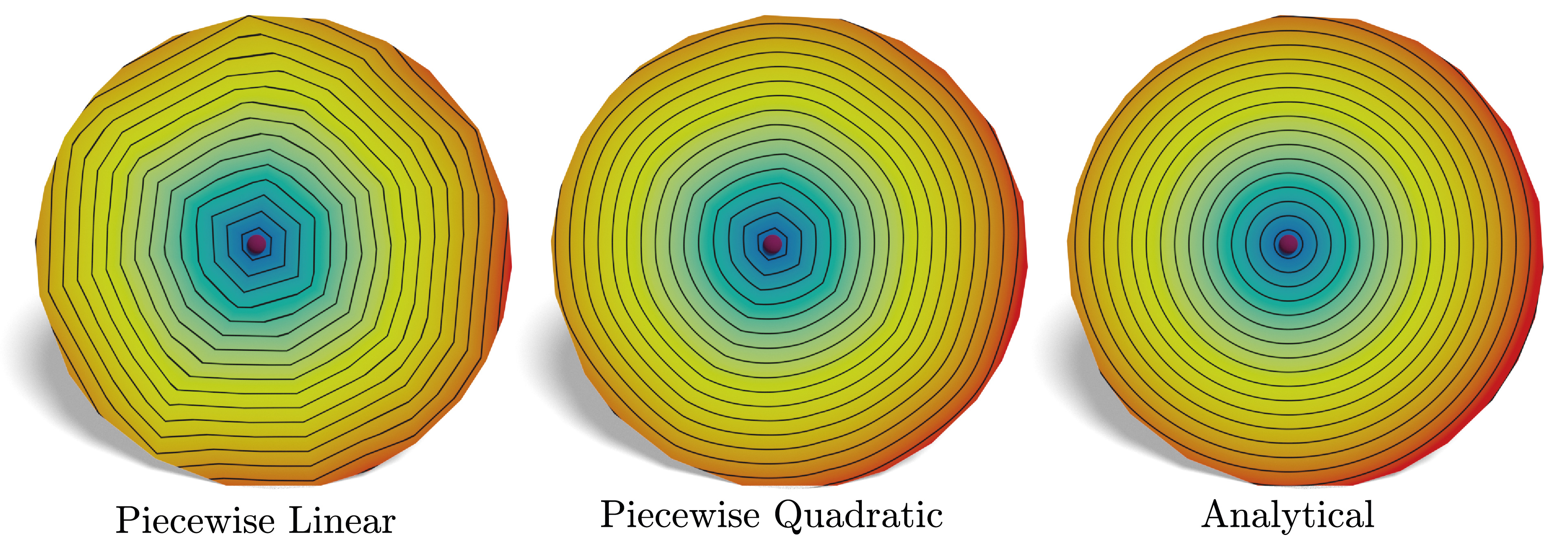}
    \caption{(left) Piecewise linear elements are not good approximators of geodsic distances near the source. (center) Our approach easily generalizes to quadratic elements. Note the improved accuracy (compare with the analytic solution (right)).}
    \label{fig:qfe}
\end{figure}

Piecewise linear elements are not good approximators of the geodesic distance near the source. Intuitively, for coarse meshing, instead of generating round isolines, PL elements lead to polygonal isolines, see e.g. the output on the disk in Fig.~\ref{fig:qfe} (left). 
Our approach generalizes to piecewise quadratic elements in a straight-forward way. 
Specifically, we replace the mass matrix, gradient and Laplacian with the corresponding matrices for quadratic elements [Boksebeld and Vaxman 2022, Appendix B]. The result is shown in Figure~\ref{fig:qfe} (center). Note that the quadratic elements lead to a better approximation (compare with the analytical solution, Fig.~\ref{fig:qfe} (right)). 

